\documentclass[manuscript,sigconf,screen]{acmart}

\usepackage[utf8]{inputenc}
\usepackage{geometry}
\usepackage{array}
\usepackage{booktabs}
\usepackage{multirow}
\usepackage{tabularx}
\usepackage{colortbl}
\usepackage{graphicx}
\usepackage{subcaption}
\usepackage{subfiles}
\usepackage{longtable}
\usepackage{threeparttable}

\usepackage{textgreek}
\usepackage{soul}       
\usepackage[normalem]{ulem}       
\usepackage{xspace}

\usepackage[most]{tcolorbox} 
\usepackage{xcolor}

\newcommand{\emoji}[1]{\includegraphics[height=1em]{emojis/#1.png}\xspace}

\newtcbox{\roundtag}{
  on line,
  colback=blue!50,
  colframe=cyan!50,
  boxrule=0.5pt,
  arc=4pt,
  boxsep=1pt,
  left=4pt, right=4pt, top=2pt, bottom=2pt,
}

\newtcbox{\fullroundtag}[1]{ 
  colback=cyan!10,
  colframe=cyan!60,
  boxrule=0.5pt,
  arc=4pt,
  boxsep=1pt,
  left=4pt, right=4pt, top=2pt, bottom=2pt,
  width=\linewidth,
  nobeforeafter,
  enhanced,
}

\newcommand{\R}{\textit{Reflexa}}
\newcommand{\Cp}{\textit{Reflection on Current Process}}
\newcommand{\Se}{\textit{Reflection on Self}}
\newcommand{\Ex}{\textit{Reflection through Experimentation}}
\newcommand{\RC}{\textit{Reflexa Core}}
\newcommand{\RD}{\textit{Reflexa Dialogue}}
\newcommand{\RL}{\textit{Reflexa Flow}}
\newcommand{\RS}{\textit{Reflexa Spark}}

\definecolor{Gray}{gray}{0.85}
\definecolor{LightCyan}{rgb}{0.88,1,1}
\newcolumntype{a}{>{\columncolor{Gray}}c}
\newcolumntype{b}{>{\columncolor{white}}c}

\definecolor{codebg}{RGB}{245,245,220}  
\definecolor{commentpurple}{RGB}{138,43,226} 
\definecolor{codered}{RGB}{255,0,0} 

\lstdefinelanguage{p5js}{ keywords={let, const, function, return, if, else, for}, keywordstyle=\color{blue}\bfseries, ndkeywords={true,false,null}, ndkeywordstyle=\color{magenta}, identifierstyle=\color{black}, sensitive=true, comment=[l]{//}, commentstyle=\color{commentpurple}\ttfamily, stringstyle=\color{codered}, morestring=[b]', morestring=[b"] }

\renewcommand\hl[1]{#1} 

\definecolor{codegray}{gray}{0.94}
\definecolor{codeblue}{rgb}{0.1,0.2,0.6}
\usepackage{tcolorbox}
\tcbset{
  highlight code/.style={
    colback=codegray,
    colframe=white,
    boxrule=0pt,
    left=3pt,
    right=3pt,
    top=2pt,
    bottom=2pt,
    boxsep=1pt,
    arc=1mm,
    enhanced,
    fontupper=\ttfamily\small\color{codeblue}
  }
}
\newtcolorbox[auto counter, number within=section]{codeblock}[2][]{colback=gray!10!white, colframe=blue!80!black, coltitle=white, fonttitle=\bfseries, sharp corners=all, top=2mm, bottom=2mm, left=2mm, right=2mm, title=#2,#1}
\AtBeginDocument{%
  \providecommand\BibTeX{{%
    \normalfont B\kern-0.5em{\scshape i\kern-0.25em b}\kern-0.8em\TeX}}}

\setcopyright{acmlicensed}
\copyrightyear{2018}
\acmYear{2018}
\acmDOI{XXXXXXX.XXXXXXX}

\acmConference[Conference acronym 'XX]{Make sure to enter the correct
  conference title from your rights confirmation emai}{June 03--05,
  2018}{Woodstock, NY}
\acmISBN{978-1-4503-XXXX-X/18/06}

\usepackage[utf8]{inputenc}
\usepackage{xcolor}
\usepackage{listings}
\usepackage{tcolorbox}
\definecolor{myyellow}{RGB}{255,242,204}
\definecolor{lightyellow}{RGB}{255,250,230}
\definecolor{codegray}{RGB}{248,248,248}
\definecolor{userrole}{RGB}{207,72,72}
\definecolor{assistantrole}{RGB}{160,32,240}
\definecolor{codepurple}{RGB}{138,43,226}
\definecolor{commentgreen}{RGB}{0,150,0}

\lstdefinelanguage{p5js}{
  keywords={let, const, function, return, if, else, for},
  keywordstyle=\color{blue}\bfseries,
  ndkeywords={true,false,null},
  ndkeywordstyle=\color{magenta},
  identifierstyle=\color{black},
  sensitive=true,
  comment=[l]{//},
  commentstyle=\color{commentgreen}\ttfamily,
  stringstyle=\color{codepurple},
  morestring=[b]',
  morestring=[b]"
}

\newtcolorbox{appendixbox}{
  colback=gray!5,
  colframe=white,
  boxrule=0pt,
  arc=4pt,
  left=6pt, right=6pt, top=6pt, bottom=6pt,
  fontupper=\small\ttfamily,
}
\begin{document}

\title[Reflexa: Reflection Scaffolding in Creative Coding]{Reflexa: Uncovering How LLM-Supported Reflection Scaffolding Reshapes Creativity in Creative Coding}

\author{Anqi Wang}
\email{awangan@connect.ust.hk}
\orcid{0000-0003-4238-6581}
\affiliation{%
  \institution{Hong Kong University of Science and Technology}
  \city{Hong Kong SAR}
  \state{}
  \country{China}
}

\author{Zhengyi Li}
\email{7802220124@csu.edu.cn}
\affiliation{%
  \institution{Central South University}
  \city{Changsha}
  \state{Hunan}
  \country{China}
}

\author{Lan Luo}
\email{lluo476@connect.hkust-gz.edu.cn}
\orcid{0009-0004-8029-2548}
\affiliation{%
  \institution{Hong Kong University of Science and Technology (Guangzhou)}
  \city{Guangzhou}
  \country{China}}

\author{Xin Tong}
\email{xint@hkust-gz.edu.cn}
\orcid{0000-0002-8037-6301}
\affiliation{%
  \institution{Hong Kong University of Science and Technology (Guangzhou)}
  \city{Guangzhou}
  \country{China}}

\author{Pan Hui}
\authornote{Corresponding author}
\orcid{0000-0001-6026-1083}
\affiliation{%
  \institution{Hong Kong University of Science and Technology (Guangzhou)}
  \city{Guangzhou}
  \country{China}
  \email{panhui@hkust-gz.edu.cn}
}
\affiliation{%
  \institution{Hong Kong University of Science and Technology}
  \city{Hong Kong SAR}
  \country{China}
  \email{panhui@ust.hk}
}

\renewcommand{\shortauthors}{Anqi Wang et al.}

\begin{abstract}
    Creative coding requires continuous translation between evolving concepts and computational artifacts, making reflection essential yet difficult to sustain. 
Creators often struggle to manage ambiguous intentions, emergent outputs, and complex code, limiting depth of exploration. 
This work examines how large language models (LLMs) can \hl{scaffold reflection not as isolated prompts, but as a system-level mechanism shaping creative regulation.} 
From formative studies with eight expert creators, we derived reflection challenges and design principles that informed Reflexa, an integrated scaffold combining dialogic guidance, visualized version navigation, and iterative suggestion pathways. 
A within-subject study with 18 participants \hl{provides an exploratory mechanism validation, showing that structured reflection patterns} mediate the link between AI interaction and creative outcomes. These reflection trajectories enhanced perceived controllability, broadened exploration, and improved originality and aesthetic quality.
Our findings advance HCI understanding of reflection from LLM-assisted creative practices, and provide design strategies for building LLM-based creative tools that support richer human–AI co-creativity. 
\end{abstract}


\begin{teaserfigure}
\centering
   \includegraphics[width=0.8\textwidth]{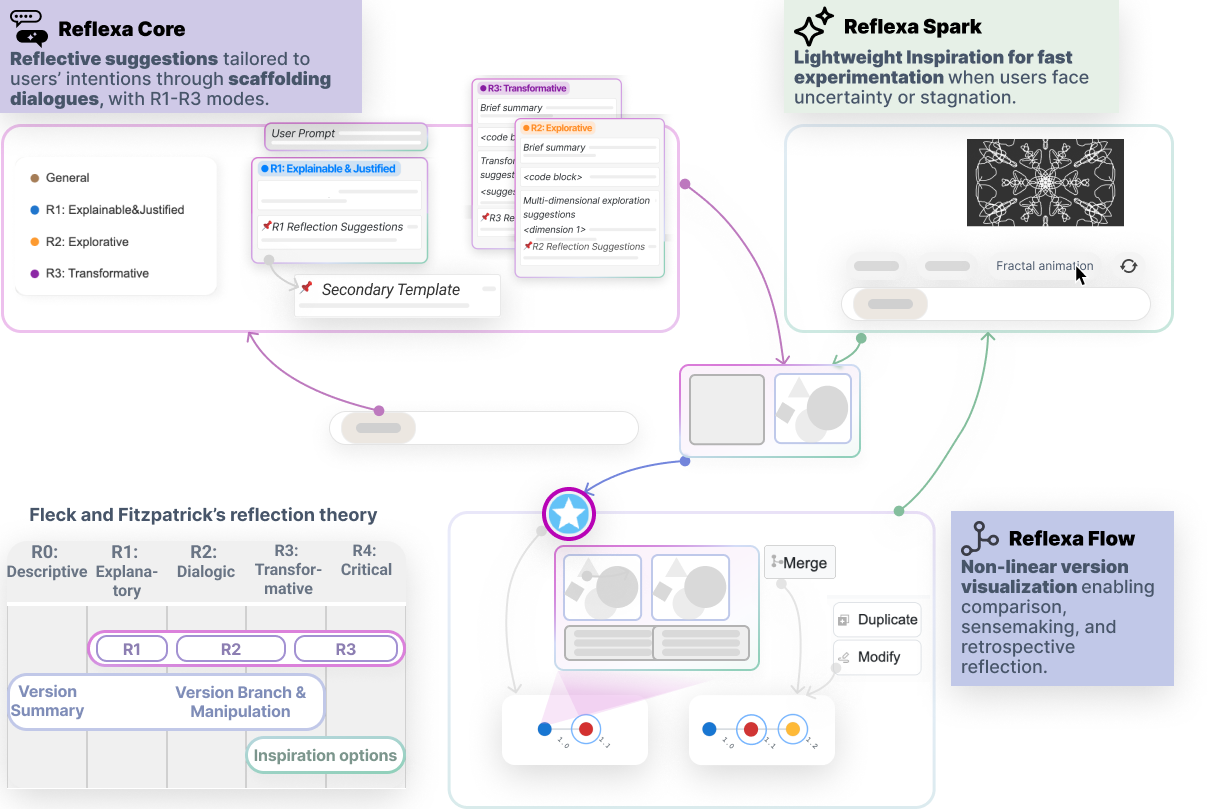}
    \caption{
    Our approach integrates three LLM-supported reflection scaffolds—Reflexa Core, which facilitates dialogic reflection through crafted templates; Flow, which structures version-based exploration; and Spark, which generates visual artifacts for direct comparison—to support reflective practice in creative coding. These scaffolds are grounded in Fleck’s reflection taxonomy (R0–R4) and Schön’s moment-to-moment reflective processes in creative practice.
    }
  \label{fig:realteaser}   
\end{teaserfigure}

\received{20 February 2007}
\received[revised]{12 March 2009}
\received[accepted]{5 June 2009}

\maketitle

\section{Introduction}
    
Creative coding is an expressive practice that merges technical construction with artistic interpretation~\cite{dufva2021coding,peppler2005creative}. Beyond generating functional code, creators engage in iterative exploration where meaning, behavior, and form co-evolve~\cite{peppler2005creative}. 
Compared with conventional programming, creative coding places particular demands: (1) repeatedly translating expressive semantic intents into lines of code~\cite{10.1145/1323688.1323689}; (2) tracking multiple variations~\cite{kery2017exploring,green1996usability}; and (3) comparing, combining, and extending emergent outputs to support further exploration~\cite{frich2019mapping,10.1145/1323688.1323689}. 
Reflection is integral to this process—not as post-hoc evaluation, but as an ongoing, situated mode of reasoning that shapes both artifact and intent, drawing on Schön’s foundational concept of \textit{``reflection-in-action''}~\cite{schon2017reflective} (reflecting in-the-moment on actions). 
These reflective cycles unfold through translating semantic intentions into computational logic, evaluating outcomes, and iteratively reformulating goals. 
However, this process places substantial cognitive demands on the creator: navigating abstract ideas, emergent outputs, and complex code structures, often within unfamiliar or underspecified creative domains~\cite{angert2023spellburst,jonsson2022cracking,wang2025pinning}. 

Recent advances in generative AI—especially large language models (LLMs), such as GPT-4o~\cite{web:chatgpt}, 
have significantly expanded opportunities for supporting reflective practices in creative coding~\cite{wang2025pinning,angert2023spellburst,jonsson2022cracking}. 
First, LLM-based tools can provide proactive, adaptive, readily available interventions in complex, coherent conversations, which can challenge assumptions and encourage deeper analysis~\cite{ford_reflection_2024,kasneci_chatgpt_2023,do_how_2022,knox_towards_2019}. 
Second, by surfacing ambiguity and offering unexpected interpretations using natural language~\cite{hamalainen2023evaluating, bang2023multitask, liu2023evaluating}, LLMs prompt artists to attribute meaning beyond mere expectation, treating unpredictability as a source of aesthetic insight and conceptual expansion~\cite{ford_reflection_2024,10.1145/3635636.3656184,10.1145/3519026,rezwana2021creative,10.1145/3490100.3516473,riche_ai-instruments_2025}. 
Despite this potential, 
one-turn dialogues often lead to premature convergence, limiting exploration and favoring predictable outputs over conceptual depth~\cite{shen_ideationweb_2025,chen_coexploreds_2025,lu_bridging_2022,riche_ai-instruments_2025}. While LLMs' unpredictability can inspire creativity~\cite{ford_reflection_2024,di2022ideamachine}, it also introduce irrelevant content that disrupts evolving goals~\cite{wang2025pinning,xu2024jamplate}. Without reflection support for evaluating ambiguity or sustaining divergence, creators risk stagnating in local optima~\cite{liu2023pretrain}, hindering rather than advancing creative intent~\cite{subramonyam_bridging_2024}. 

Few empirical studies explore how LLM-based systems can prompt, structure, or sustain reflective thinking during creative coding (e.g.,~\cite{wang2025pinning}). 
Following reflection theory~\cite{baumer2015reflective,fleck_designing_2012}, recent systems support reflection scaffolding for rich situations in multi-turn dialogues using graphical tools, navigating ambiguity and evolving ideas~\cite{xu2024jamplate,xu_productive_2025}. 
However, such text-based dialogue breaks away from improvisational creative practices. 
Other creativity support works adapt visual approaches, such as node-based representations to manipulate creative variations—refining, branching, merging versions~\cite{riche_ai-instruments_2025,angert2023spellburst,subramonyam_bridging_2024,zhou2024nonlinear,riche_ai-instruments_2025}, or employ visual generative artifacts to foster creative iteration, they rarely adapt reflection scaffolding to the creative process. 
Current progress underscores a gap that needs to incorporate multi-level reflection scaffolding approaches (e.g., dialogue, visual artifacts) to support creative coding~\cite{jonsson2022cracking}, thus motivates our research questions: 

\begin{itemize}
\item [\textbf{RQ1.}] 
    \textit{\hl{How can an LLM-based system be designed to scaffold reflection during creative coding?}}
\item [\textbf{RQ2.}] 
    \textit{How does such \hl{LLM-supported reflection scaffolding} shape reflective behaviors, creative processes, and resulting creative outcomes?}
\end{itemize}
This paper proposes a practical reflection scaffolding with LLMs in creative coding. 
We conducted formative studies and identify three design guidelines that highlights needs for multi-level reflection scaffolding in the creative coding. Drawing on reflection theories and technologies (i.e.,~\cite{baumer2014reviewing,baumer2015reflective,cho_reflection_2022,macneil_framing_2021,xu2024jamplate}), we iteratively developed \textit{Reflexa}, enables users to (1) engage in multi-level dialogic reflection with LLMs (\textit{Core} in Figure~\ref{fig:realteaser})
(2) employ node-based versions to revisit and navigate creative variation (\textit{Flow} in Figure~\ref{fig:realteaser})
, and 
(3) generate visual artifacts for exploration (\textit{Spark} in Figure~\ref{fig:realteaser}). 
As intuitive creative practices, \textit{Reflexa} were be conceptualized reflection as a constellation of interrelated processes rather than as a singular event. 
We evaluated this reflection scaffolding through a within-subjective study with 18 participants. Our findings reveals how the reflection scaffolding elicits reflective behavior, and effects as pivotal mediator between interaction and creative outcomes, by leveraging interplays of the \textit{Reflexa} features.

Our contribution is thus threefold. 
\hl{First, we identify challenges and develop design guidelines for scaffolding reflection during creative coding with AI interaction.} 
\hl{Second, we develop a system-\textit{Reflexa}, that integrates multiple reflection scaffolds within a unified interaction paradigm, built upon Fleck and Fitzpatrick's reflection taxonomy.} 
\hl{Third, we extend the concept of reflection to the human-AI co-creation context by validating \textit{Reflexa} with a user study, thus reframing reflection as system-mediated. 
} %

\section{RELATED WORK}
    
\subsection{Supporting Reflection in Creative Process}~\label{sec:rw-reflection}
    The term ``reflection'' encompasses a broad range of concepts, including ``conscious, purposeful thought directed at a problem to understand it and form integrated conceptual structures''~\cite{sedig2001role}, and a process ``in which people recapture their experience, think about it, mull it over, and evaluate it''~\cite{bateman2012search}. 
    Reflection has long been investigated in the HCI field as a fundamental outcome in technology design~\cite{sengers2005reflective,sharmin_reflectionspace_2013}. 
    Scholars address reflection among technology users, explicitly exploring how reflection can enhance user experiences. Fleck and Fitzpatrick, for example, identify five levels of increasing reflectivity~\cite{fleck_reflecting_2010}: 
    \emph{R0}, descriptive reflection, involves revisiting events without further elaboration; 
    \emph{R1}, explanatory reflection, adds justifications or reasons for actions; 
    \emph{R2}, dialogic reflection, explores relationships and alternative perspectives; 
    \emph{R3}, transformative reflection, challenges personal assumptions and reorganizes   understanding; and 
    \emph{R4}, critical reflection, considers broader social, ethical, and contextual implications. 
    Similarly, Baumer et al. critically distinguish reflection types as \textit{breakdown, inquiry,} and \textit{transformation}~\cite{baumer2014reviewing, baumer2015reflective}. These frameworks provide a theory foundation for understanding the design dimensions of technologies intended to promote reflection~\cite{baumer2015reflective}. 

\subsubsection{Reflection in Creative Process}
    The identification of problem-solving paths in the complex, creative design process—often hindered by ill-defined, incomplete, conflicting, and changing requirements~\cite{buchnan1992wicked}—requires action-oriented reflection. This involves contemplating what occurred and how it unfolded, followed by adapting future actions based on the insights gathered~\cite{liedtka2018design}. This intuitive reflection process uncovers tacit knowledge derived from practical behaviors, which constitutes an essential component of design activities~\cite{schoormann_designing_2022, Schon1992designing}. The main focus is on the integration of reflective practices in the design process~\cite{ghajargar_thinking_2018,guillaumier2016reflection,candy_creative_2020}, allowing the creators to identify unconscious aspects of the creative problem and rethink dominant creative choices.

    In creative context, reflection has found widespread use throughout the entire creative process~\cite{hummels2009reflective, mendels2011freed, sharmin_reflectionspace_2013, yoo2013value, dalsgaard2012supporting, hansen2012productive}, spanning from defining the problem~\cite{dijk2011noot} to refining outcomes~\cite{hailpern2007team}. 
    Schön’s concept of reflective practice-``reflection-on-action''~\cite{schon2017reflective}-described as a reflective ``conversation with the materials'' of design situation~\cite{Schon1992designing}-has been applied to improve the quality of interaction design~\cite{lowgren2004thoughtful}. 
    Several studies have focused on various approaches to eliciting reflection and designing interactive system for creative process~\cite{baumer2014reviewing, bentvelzen2022revisiting,brade2012ontosketch}.     
    \hl{Whereas such refection underexplores explicit definition and lacks effective measurements. Ford et al. addressed this gap by introducing the self-reported \textit{Reflection in Creative Experience} (RiCE) questionnaire~\mbox{\cite{ford_reflection_2024}}, which captures moment-level refection engagement during creative activity. 
    RiCE developed three dimensions grounded in established reflection theories: reflection on process (Cp), self (Se), and experimentation (Ex). 
    Cp synthesizes Schön’s reflection-on-action and captures how creators interpret system behaviour as it unfolds; 
    Ex represents hypothesis-testing and variation cycles characteristic of creative cognition and aligns with reflection-in-action and Fleck and Fitzpatrick’s R2–R3 exploratory reflection; and 
    Se captures the reframing of goals and intentions central to Fleck’s meta- or higher-level abstraction (R1, R3)~\mbox{\cite{ford_reflection_2024}}.
    This framework advances prior work by conceptualizing reflection not as a purely retrospective act but as a recurring, moment-to-moment process embedded within creative action. 
    }
These works offer a theoretical lens for understanding and designing reflection in creative processes.



    \subsubsection{LLM Capabilities and Tools for Eliciting Reflection}
HCI research has examined how AI tools facilitate reflection in creative practices. Jonsson~\cite{jonsson2022cracking} highlighted AI’s role in understanding code creation and supporting reflection-in-action~\cite{schon1987educating}. Scholars have mapped reflection taxonomies~\cite{baumer2014reviewing, fleck_reflecting_2010} onto system features, showing how specific affordances elicit distinct reflective behaviors. Wang~\cite{wang2025pinning} further demonstrated that different LLM prompting strategies produce varying frequencies of reflection in creative coding, though practical scaffolding remains limited. Current LLM systems (e.g.,~\cite{druga_scratch_2023, jonsson2022cracking, wang2025pinning}) show potential in supporting reflection and divergent thinking through several strategies:

\textit{Generative Artifacts for Reflection.} By grounding projects in LLM-generated outputs, artists can evaluate constraints, anticipate outcomes, and iteratively refine intentions~\cite{schon1987educating}. LLMs also facilitate semantic leaps via natural language, enabling expansive exploration~\cite{angert2023spellburst}. Open-ended prompts, including randomized or ambiguous suggestions, support reflection from revisiting past ideas to stimulating transformative thinking~\cite{cho_reflection_2022,macneil_framing_2021,sharmin_reflectionspace_2013,baumer2015reflective,fleck_designing_2012}.

\textit{Dialogic Reflection Scaffolding.} LLM-based conversations provide adaptive, immediate interventions, engaging users in coherent dialogues that challenge assumptions and encourage deeper analysis~\cite{kasneci_chatgpt_2023,do_how_2022,knox_towards_2019,karinshak_working_2023,10.1145/3635636.3656184,10.1145/3519026,rezwana2021creative,10.1145/3490100.3516473}. By posing questions rather than answers, LLMs foster critical inquiry, helping designers examine assumptions and broaden perspectives~\cite{danry2023dont,lee2024conversationalagentscatalystscritical,doi:10.1177/07356331221077901}.

\textit{Visually Node-based Version Exploration.} Integrating LLMs with node-based, multi-version visual interfaces supports the organization of scattered ideas and exploratory variations~\cite{subramonyam_bridging_2024,zhou2024nonlinear,angert2023spellburst,riche_ai-instruments_2025}. Tools like Spellburst enable ``revisiting reflection'' via version navigation and visual feedback, enhancing nonlinear, exploratory creative coding~\cite{angert2023spellburst}. LLMs can further suggest conceptual connections and references, fostering integrated reflective processes~\cite{chang2023promptart}.

Recent systems, such as Jamplate~\cite{xu2024jamplate}, explicitly scaffold multi-level reflection in LLM-driven dialogues based on reflection theories~\cite{fleck_reflecting_2010}, aiding users in navigating ambiguity and refining ideas. AI-Instruments~\cite{riche_ai-instruments_2025} extends this with graphical prompt iteration and visual feedback. Nonetheless, most tools lack systematic analysis of how prompting or dialogue strategies shape reflective processes and their creative outcomes.

\subsection{LLM Tools for Creative Coding}
    \textit{Creative coding} refers to a branch of programming that prioritizes artistic and expressive outputs, subordinating functionality to creative expression~\cite{peppler2005creative,shiffman2012nature,ackermann_programming_2020}. Generative art represents a specific approach to creative coding in which artists “program computers to undertake creative instructions”~\cite{tempel2017generative,ward1999how}. In this process, artists explore and experiment with code to generate algorithmic variations and outputs~\cite{kery2017exploring,kery2017variolite}. Through such practices, they frequently engage with phenomena such as emergence, randomness, and interaction to produce novel and unexpected creative results~\cite{phillips2011algorists,creators2012case,creators2012case,ward1999how}.
    Although the needs of artists engaged in creative coding overlap with those of traditional programmers, they also require specific affordances that support rapid iteration and semantic exploration. ~\citet{angert2023spellburst} identified three unique needs in exploratory creative coding: (1) repeatedly translating expressive semantic intents into lines of code, (2) tracking and managing multiple creative variations, and (3) comparing, combining, and extending emergent outputs to fuel further exploration.
    
    Recent advancements in large language models (LLMs)—such as GPT-4o, Gopher~\cite{rae2021scaling}, and LaMDA~\cite{thoppilan2022lamda}—have demonstrated strong capabilities in natural language understanding and programming-related tasks~\cite{liu2023evaluating,hamalainen2023evaluating,bang2023multitask}. These models can generate coherent, high-quality code from textual prompts~\cite{chakraborty2022natgen,dakhel2023github,fried2022incoder}, reducing the technical burden and enabling broader participation in creative coding. Interestingly, even the hallucinations produced by LLMs—unintended or imprecise outputs—can provoke novel insights and spark creative reflection~\cite{wang2025pinning,jonsson2022cracking}, making them not only facilitators of automation but also sources of inspiration.

    Prior research has examined how LLM-based systems support exploratory creative coding by either generating content or providing interactive programming interfaces. For instance, Spellburst~\cite{angert2023spellburst} introduced a visual node-based tool that facilitates code generation and variation exploration through semantic programming. It includes several key features: (1) version control systems to support history tracking~\cite{10.1145/1323688.1323689}; (2) a closer mapping between semantic concepts and syntactic representations, improving usability in creativity support tools~\cite{kery2017exploring,green1996usability}; and (3) AI-augmented features that foster creative exploration~\cite{frich2019mapping,10.1145/1323688.1323689}.

    Given the open-ended and nonlinear nature of creative coding~\cite{jonsson2022cracking}, supporting such practices requires scaffolds that adapt to evolving intentions, respond to context, and surface meaningful semantic cues. These gaps underscore the need for adaptable strategies in LLM-supported creative coding—approaches that go beyond singular reflective moments to support continuous, multi-layered reflective engagement. 
    Thus, our works incorporates integrating strategies to elicit reflection during exploratory creative coding, by (1) creative generation for reflection-in-action; (2) dialogic question-answer using multiple-level reflection supports; (3) visual tracking and manipulation of historical variations for further exploration. 
    Accordingly, our work integrates targeted mechanisms to elicit reflection throughout exploratory creative coding:
(1) generation-based support for reflection-in-action;
(2) dialogic prompts structured across multiple levels of reflection; and
(3) visual tracking and manipulation of historical variations to foster ongoing exploration.



\section{Formative Study}

\label{sec:formativestudy}

\begin{figure*}[h]
    \centering
    \includegraphics[width=\linewidth]{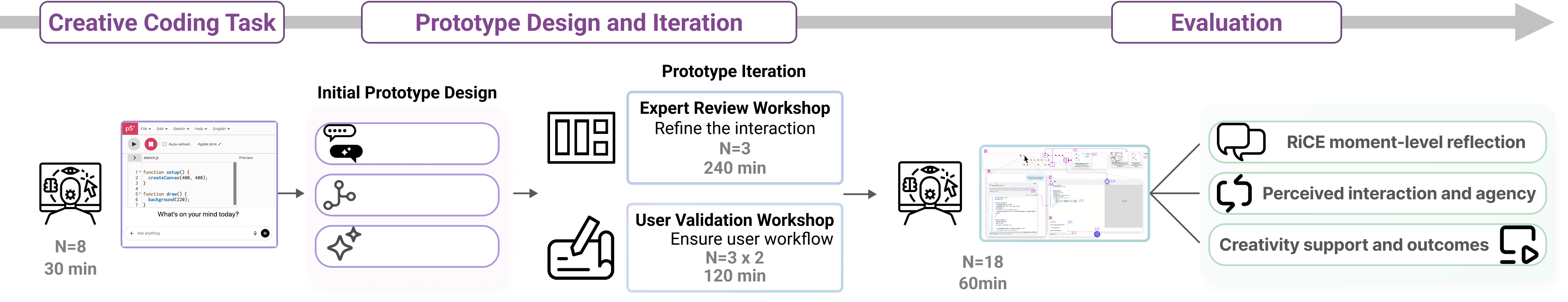}
    \caption{\hl{Study procedure across three stages.}}
    \label{fig:fs-studyprocedure}
\end{figure*}

\hl{To examine the reflection practices and challenges of LLM-assisted creative coding and inform system design, we conducted a formative study with domain users who have creative coding experiences. The study included (1) understanding reflection by interaction analysis, (2) prototype design, informed by design guidelines in the first study, and a follow-up expert feedback (3) prototype iteration. Figure~\mbox{\ref{fig:fs-studyprocedure}} illustrates our study procedure including formative study and evaluation (Section~\mbox{\ref{sec:userstudy}}).} 

\subsection{Understanding Reflection in Creative Coding}
\subsubsection{Study Design}
To understanding challenges and patterns of reflection during creative coding, we conducted a creative coding workshop with eight artists. 
This study also builds on prior work examining when, what, and how reflection occurs in LLM-based dialogues~\cite{wang2025pinning}. 

\textit{Participants.} 
We recruited eight participants (three male, five female; ages 20--25) from a creative coding workshop at a fine arts academy. The workshop targeted students in their final year or above within relevant disciplines, ensuring they possessed at least three years' experience in creative programming. Pre-workshop questionnaires indicated all participants had over one year's experience using large language model tools such as ChatGPT. Participants were initially randomly organized into small workshop groups, with one representative from each group randomly selected, yielding a stratified random sample. Recruitment was conducted on-site, and all participants provided informed consent. Participants take part voluntarily without compensation. The study protocol was approved by the university IRB.

\textit{Task and Procedure.}
Each participant completed a creative coding task requiring the design of a visual or interactive artifact using a combination of self-authored prompts and LLM-generated suggestions. Tasks emphasized open-ended exploration, encouraging iterative refinement toward functional and expressive goals. 
Data collection included screen recordings, conversation logs with the LLM, coding results, annotated design artifacts, and post-task interviews with participants’ consent for audio recording. 

\textit{Data Analysis.} 
\hl{We analyzed participants’ interaction histories with LLMs using a thematic analysis~\mbox{\cite{Braun01012006}} to identify reflective patterns and challenges. Two authors initially open coded observed behaviors and prompts. These codes were then grouped into themes, which were carefully reviewed and discussed to identify the key findings of the study. We coded each interaction round in: (1) user intent, (2) user behaviors, involving \textit{rethinking, questioning, or revising ideas}, drawing on prior literature~\mbox{\cite{wang2025pinning}}, (3) occurred LLM responses. Then we defined three stages in creative coding, including ideation, function, and feature, according to user intent and behaviors. Using these codes, we finally inductively identified (1) reflection practices, mapping them to Fleck's reflection level~\mbox{\cite{fleck_reflecting_2010}}; (2) reflection challenges.} 


\setulcolor{blue} 
\setul{0.5ex}{0.3ex} 
\definecolor{tab:blue}{RGB}{31,119,180}
\definecolor{tab:yellow}{RGB}{255,219,0}
\definecolor{tab:green}{RGB}{140,220,100}

\begin{figure*}[h]
    \centering
    \includegraphics[width=0.99\textwidth]{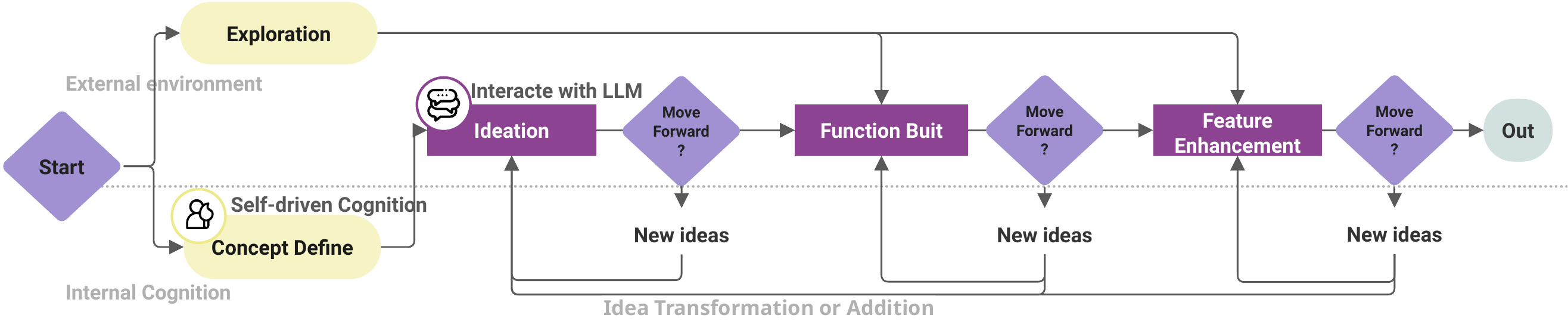} 
\caption
{\hl{The creative coding process with LLMs unfolds through three stages: ideation, function built, and feature refinement.}
}
    \label{fig:FS-workflow}
\end{figure*}

\subsubsection{Reflection Practices and Challenges in LLM-Assisted Creative Coding}
~\label{sec:reflectionpractices}
~\label{sec:reflectionchallenge} 
Our analysis demonstrated creative coding is a non-linear exploration process across Ideation, Function, and Feature stages, we illustrated such process in Figure~\ref{fig:FS-workflow}. 
We revealed four recurring patterns across the whole process: 

\textit{(1) Troubleshooting and Goal Clarification.} Participants often adjusted prompts or code when outputs diverged from expectations, demonstrating the need to scaffold transitions from low-level debugging to higher-order reflection. This stage highlights the challenge of \textbf{unclear creative expression (C1), as participants struggled to articulate intentions clearly to the LLM.}

\textit{(2) Feature Refinement and Exploration.} Once functionality stabilized, participants evaluated alternatives and extended features, such as mapping rhythm to visual tension and making deliberate choices to achieve desired expressive outcomes. This stage reveals the challenge of \textbf{limited prediction and evaluation (C2), since participants rarely analyzed how current outputs aligned with prior goals.}

\textit{(3) Goal Reframing.} When progress stalled or goals were unclear, participants reset directions or reconsidered outcomes. This stage illustrates the challenge of \textbf{decomposing abstract ideas (C3), as metaphorical or emotionally oriented concepts were difficult to translate into implementable code.}

\textit{(4) Nonlinear Iteration and Selective Reuse.} Participants selectively reused prior code, highlighting the need for structured support to integrate past work. This stage underscores the challenge of \textbf{ineffective reflection in trial-and-error (C4), since users lacked mechanisms to assess the significance of incremental adjustments.} 
This stage-oriented view shows how reflective practices expose specific challenges, emphasizing the need for scaffolds that support goal articulation, iterative evaluation, decomposition of abstract ideas, and structured exploration.

\subsubsection{Design Goals}
Our findings lead to the following three design goals:
First, such a framework needs to challenge the single original creativity of artists through engaging designers and artifacts in co-creative dialogues-aligning with prior studies on reflection supported system~\cite{sharmin_reflectionspace_2013,ford_reflection_2024,jonsson2022cracking}, which proposes \textbf{DG1) scaffolding intent articulation via context-aware prompts}. (C1, C2, C3)

Second, in line with prior findings on supporting nonlinear creative process~\cite{zhou2024nonlinear,angert2023spellburst,qian_shape-it_2024,suh_sensecape_2023,wang_survey_2025,qin_toward_2025,ford_reflection_2024,li_what_2021}, such a framework should incorporate interactive operations for visual tracking history/versions to navigate their creative processes, as well as iterate previous prompts and outputs. This raise goals of \textbf{DG2) supporting reflection practices throughout the non-linear process to facilitate divergent-convergent thinking cycles.} (C4)

    

Third, such a framework also needs to main relatively level reflection during process. Given we found breakdown occurred frequently (e.g., repeating same instruction or organizing wording), \textbf{DG3) turnning low-level reflection (breakdowns) into high-level reflective opportunities is necessary.} 
This approach helps maintain a positive user experience in the face of setbacks and encourages reflective practices that enhance programming skills and user confidence. (C1, C3)
We elaborate several functions and detail in Section~\ref{sec:systemdesign}. 


\subsection{Prototype Design}
\hl{Building on above identified reflective needs of creative coders and three DGs, we developed prototype through an iterative design process involving theoretical grounding, expert-driven interaction refinement, and user workflow validation.} 
\subsubsection{Theoretical integration} To establish a structured reflective foundation, we synthesized Schön’s reflection-in/on-action~\mbox{\cite{Schon1992designing, schon2017reflective}} with Fleck and Fitzpatrick’s levels of reflection~\mbox{\cite{fleck_reflecting_2010}}, alongside established reflection frameworks in creative practice~\mbox{\cite{candy_creative_2020, moon1999reflection, bain_using_1999}}. 
\hl{We integrated these theories by aligning their complementary emphases—using Schön’s descriptive grounding of reflection-in-action (e.g., framing a situation, taking action, and interpreting its effects), Fleck’s structured levels of reflection, and creative-practice consideration of exploratory and transformative shifts. 
We also drew inspiration from LLM-supported creativity systems (e.g.,~\mbox{\cite{xu2024jamplate,xu_productive_2025}}).} 
\hl{This conceptual integration allowed us to articulate reflection as a progression from descriptive articulation to dialogic exploration and transformative reframing.} 

\begin{figure*}[h]
    \centering
    \includegraphics[width=0.75\linewidth]{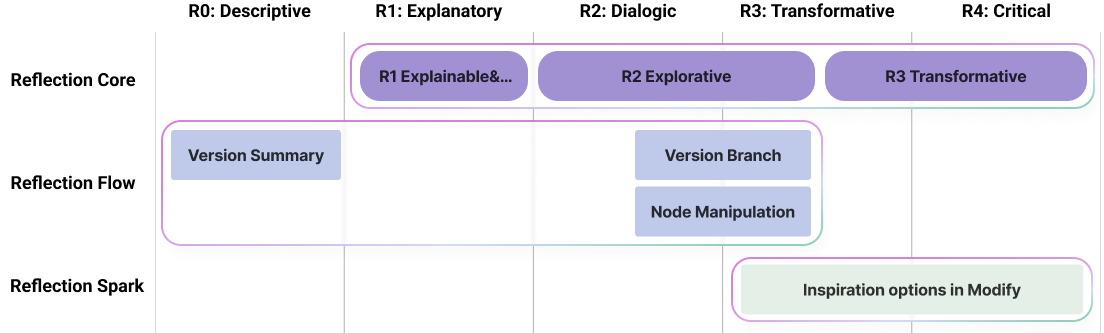}
    \caption{\hl{Overview of Reflexa system features grounded on Fleck \& Fitzpatrick's reflection framework (R0-R4)}~\cite{fleck_reflecting_2010} (see Section~\ref{sec:rw-reflection}). 
    }
    \label{fig:feature-theory}
\end{figure*}

\subsubsection{Initial Prototype Design}
\hl{Guided by this theoretical mapping, we developed a mock-up prototype that contains two-tier reflective prompt structure and three design modules (Figure~\mbox{\ref{fig:feature-theory}}). First, we operationalized reflection as three dialogic modes—descriptive/justificatory (R1), explorative (R2), and transformative (R3)—corresponding to users’ needs to articulate intentions (C1), examine alternatives (C2–C3), and reframe goals (C4).
} 
\hl{Second, we translated these modes into interaction scaffolds, including structured question templates, example-driven cues, and alternative-perspective prompts inspired by peer dialogue and design critique practices~\mbox{\cite{hubbard2023dimensions, xu2024jamplate}}.} 
Third, drawing on DG1 context-aware prompts, we elaborately designed several dialogic prompts as secondary templates under three modes (R1-R3) grounded on needs of reflection-on-design. This process involves remapping the previously codebook (Table~\mbox{\ref{tab:reflection_codebook_enriched}}), with multiple modes corresponding to different reflection needs. 


\begin{table*}[h]
\centering
\small
\caption{\hl{Codebook of representative user prompts, their reflection focus, and corresponding system needs informing the seven Reflexa reflection secondary templates (illustrated in Section~\mbox{\ref{sec:system-reflexacore}}).
}}
\begin{tabular}{p{4cm} p{3.4cm} p{3.4cm} p{3.6cm}}
\textbf{Quote} & \textbf{Reflection Focus} & \textbf{Reflection Need} & \textbf{Specific Design(s) in \R{}} \\
\hline
\textit{``...the animation still doesn’t match what I imagined. Can you explain which parts...?''} (P3) 
& Diagnosing mismatches between intention and system behavior; clarifying functional meaning
& (a) Support articulation of implicit intentions behind code-level adjustments 
& \textbf{R1–Design Detail Justification}, \textbf{R2–Module–Experience Relations} \\
\hline
\textit{``I think I can add two more steps to make it more dynamic... Extend interaction modality from mouse click to drag''} (P1)
& Expanding interaction logic to enrich experiential dynamics
& (b) Scaffold anticipatory reasoning about consequences of interaction changes
& \textbf{R2–Module–Experience Relations}, \textbf{R3–Transform Creative Direction}
\\
\hline
\textit{``Forget those special effects, I’ll just make something new''} (P5) 
& Reframing creative direction after breakdowns or dead-ends
& (c) Enable conceptual reframing and redirection at creative turning points
& \textbf{R1–Creative Motivation}, \textbf{R3–Transform Creative Direction}
\\
\hline
\textit{``I know where the problem is and I’m considering whether to use the earlier method again.''} (P8) 
& Evaluating prior versions and comparing alternative approaches
& (b) Support structured comparison across versions to guide experiential reasoning
& \textbf{R2–Conceptual Connections}, \textbf{R3–Shift Visual Style} \\
\end{tabular}
\label{tab:reflection_codebook_enriched}
\end{table*}

\hl{We then developed low- and mid-fidelity prototypes, \textit{Reflexa}, in FigJam and Figma, establishing three foundational modules: the dialogue-oriented \mbox{\RC{}}, the version-based \mbox{\RL{}}, and the inspiration-oriented \mbox{\RS{}}—each aligned with different forms of reflective engagement. 
} 

\subsubsection{Expert Interaction-Refinement Workshop}


%

\hl{To translate the initial prototype into an implementation-ready model, we conducted workshops to refine interaction patterns with three experts.} 

\textit{Participants.} We recruited three professional product designers with 3-5 years creative-tool design experience and \hl{prior creative coding experience} via social media. \hl{We screened for at least 1 years creative-coding and at least 1 year LLM-assisted practice.} \hl{Recruitment occurred through university posters, volunteered without compensation, and provided informed consent. The study protocol was approved by the university IRB.} 

\textit{Procedure.} \hl{The workshop followed established expert-review and scenario-based evaluation protocols~\mbox{\cite{nielsen1990heuristic,10.5555/189200.189214}}, and began with introducing study objective and procedure, and signed consent forms. This workshop had two sessions. 
\textit{Session 1: Component Evaluation (2h).}
We conducted: (1) \textit{problem framing}: reviewing reflection challenges and identifying interface-level opportunities grounded in the three modules;
(2) \textit{prototype walkthrough} of three modules via the Figma prototype; (3) \textit{heuristic evaluation} using Nielsen's heuristics~\mbox{\cite{nielsen1990heuristic}}, assessing interaction coherence, affordance effectiveness, and Study 1 alignment (C1–C4), documented via think-aloud in a shared Figma board. (4) \textit{collaborative critique} using affinity diagramming to prioritize 3–5 critical issues (version navigation, code-reflection linking, automation-control balance). 
\textit{Session 2: Refinement (2h).}
Participants completed: 
(1) \textit{interaction refinement} addressing Session 1 priorities via Figma annotations and design patterns; (2) \textit{design consolidation} producing annotated Figma specifications for implementation.} 

\textit{Data Analysis.}
\hl{Sessions and artefacts (screenshots, Figma boards, annotations) were recorded. Recordings were initially autotranscribed and subsequently manually verified for accuracy. We employed the inductive thematic analysis method~\mbox{\cite{braun2012thematic}} to identify recurring design patterns and usability concerns. 
Two researchers, including a facilitator and an assistant who participated in all workshops, – independently coded transcripts and annotated visual materials, then met iteratively to discuss discrepancies and refine codes until consensus was reached. This method was chosen for its suitability in exploratory design contexts where the goal is to extract emergent patterns from qualitative expert feedback~\mbox{\cite{braun2012thematic}}.}

\subsubsection{Design Refinements.}
\hl{We adjusted the prototype according to yielded three core refinements: (1) streamline reflection pathways by simplifying R1–R3 choices, and embedding template prompts directly into dialogue, rather than card options; (2) reduce interface load through collapsible summaries, a coding-first layout, and removal of nonessential search/filter functions; and (3) restructure subtasks into a clearer nonlinear node-version model to support coherent reflective progression.} 

\subsection{Prototype Iteration}
\subsubsection{User Workflow-Validation Workshops}~\label{sec:study2:codesign}
\hl{While expert workshops ensured interaction design quality and technical feasibility, user validation was necessary to assess whether the refined prototypes aligned with actual creative coding workflows and reflective practices in realistic use contexts.} 
Thus, we conducted two co-design workshops (3 participants each, 60–80 minutes) focusing on (1) \textit{Function Prototype Iteration} and \textit{User Journey Iteration}. 

\textit{Participants.}
\hl{We recruited six creative coders (4 male, 2 female, aged 22–28) with 2–5 years of creative coding experience across diverse backgrounds: creative media art (n=3), digital media art (n=2), and generative art (n=1). All had prior experience with LLM-assisted coding tools (e.g., GitHub Copilot, ChatGPT). Recruitment occurred through creative coding communities and university design programs. Participants received no compensation and volunteered with informed consent. The study protocol was approved by the university IRB.} 

\textit{Procedure.}
\hl{The workshops began with an overview of its objectives and procedures, followed by the collection of informed consent. Participants were then provided access to a Figma board containing both mid- and high-fidelity prototypes. Each session comprised: 
    (1) \textit{Mid-fidelity prototype review}: A researcher introduced three modules of Reflexa (\mbox{\RC{}}, \textit{Flow}, \textit{Spark}) outlined their respective purposes and features. Participants evaluated each module’s intended function, reflective affordances, and potential breakdowns to determine the conceptual coherence of the system relative to their creative-coding practices; 
    (2) \textit{User-journey walkthrough}: Using a high-fidelity prototype, the researcher demonstrated a step-by-step creative-coding scenario (e.g., prompting, system responses, version branching, inspiration cues). Participants discussed conditions under which they would versus did engage each reflective module; 
    (3) \textit{Collaborative reflective discussion}: Participants examined the alignment between the reflection scaffolds and their actual reflective needs and proposed refinements to prompt scaffolds, versioning workflows, and inspiration mechanisms. 
Participants were encouraged to think aloud throughout the sessions. All workshops were audio-recorded with consent.}

\textit{Data Analysis.}
\hl{Recordings were transcribed verbatim. Using the same thematic analysis approach~\mbox{\cite{braun2012thematic}}, two researchers independently coded critical 
incidents related to: (1) feature discoverability, (2) reflection 
mode alignment, and (3) customization requests. Themes were 
synthesized through affinity diagramming. Codes were discussed iteratively until consensus, then synthesized into themes through affinity diagramming. This consistent analytical approach enabled direct comparison with expert workshop findings while capturing situated user feedback~\mbox{\cite{braun2012thematic}}.} 

\subsubsection{Design Outcomes}
Workshops yielded four key improvements: 
(1) direct visual feedback for \textit{Version Navigation} and \textit{Reflexa Spark}; 
(2) a more challenging response from R3 mode. 
(3) user-driven customization in version description and management (add/delete). These specifications informed the final implementation (Section~\ref{sec:systemdesign}). 




\section{System Design}
    \label{sec:systemdesign} 
\begin{figure*}[h]
   \includegraphics[width=\textwidth]{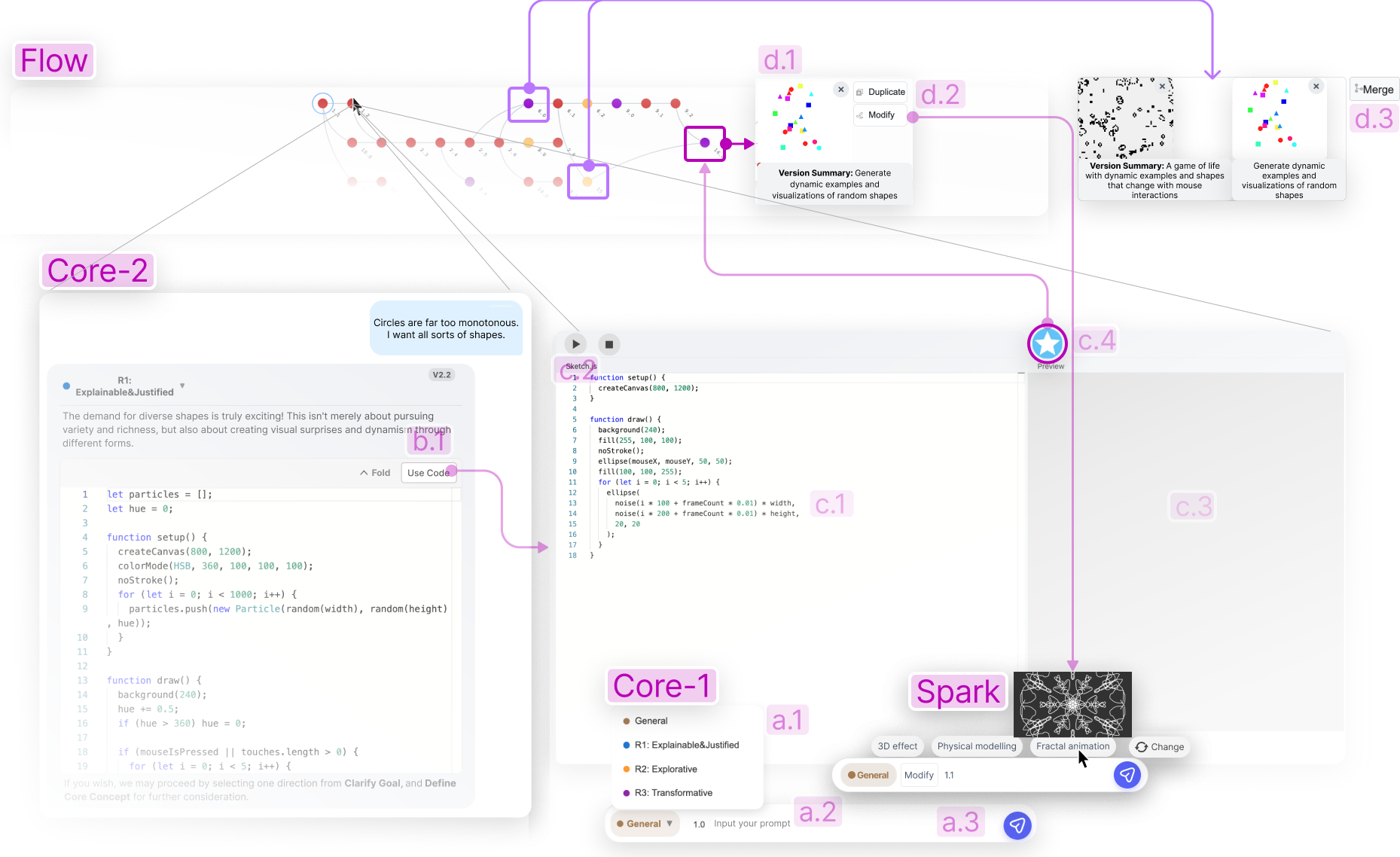}
     \caption{
        The Reflexa interface is an LLM-based creative coding tool supporting reflection. Users begin in the chatbox of \RC{}, select a reflection-supported mode R1-R3 (a.1), enter a prompt (a.2), and click ``send'' (a.3) to receive generated code with reflective suggestions in the dialogue area of \RC{}. 
        This code can be copied (b.1) into the editor (c.1), executed (c.2), and previewed in the code editor area (c.3).  
        The ``collect'' (c.4) button saves code as nodes in the \RL{}, each supporting a visual preview (d.1), modification (d.2), or merging (d.3) of two versions, which can initiate quick iteration in \textit{Reflexa Spark} above chatbox. Selecting a node synchronizes the dialogue panel and p5.js editor with the chosen version. Spark entered by modify in the Flow.
    }
  \Description{}
  \label{fig:teaser}   
\end{figure*}

Based on our three DGs and prototyping iterations, we designed and implemented \R{}, an LLM-based reflection support system to assist artists in creative coding. In this section, we elaborate 1) the three-feature interface of \textit{Reflexa}; and 2) its LLM back-end and technical implementation. 

\subsection{Reflexa Interaction Design}  
\textit{Reflexa} is embedded in \texttt{p5.js} API to integrate LLMs into the coding workflow (Figure~\ref{fig:teaser}), responding to reflection needs identified in our formative study (Section~\ref{sec:formativestudy}), including three intertwined reflection strategies. The system comprises three interdependent panels corresponding to distinct types of reflective interaction: 

\subsubsection{Reflexa Core: Reflection Mode \& Template Scaffold Selection during Dialogue}~\label{sec:system-reflexacore} 

\begin{figure*}[h]
    \centering
    \includegraphics[width=0.7\linewidth]{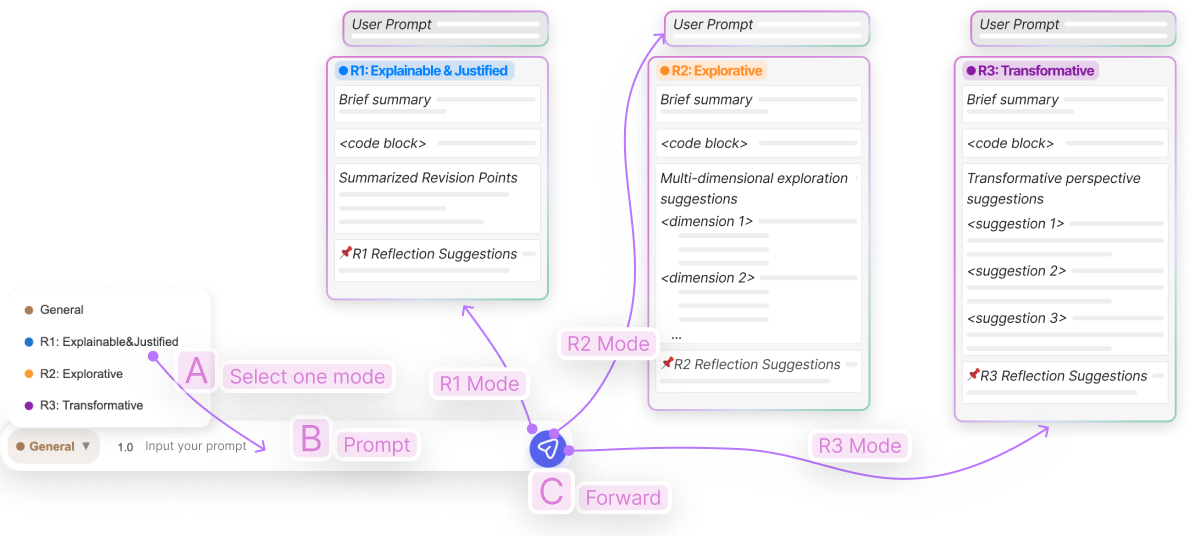}
    \caption{Interaction flow in Reflexa Core with three modes.}
    \label{fig:3Rs-template}
\end{figure*}

\begin{figure*}[h]
    \centering
    \includegraphics[width=0.6\linewidth]{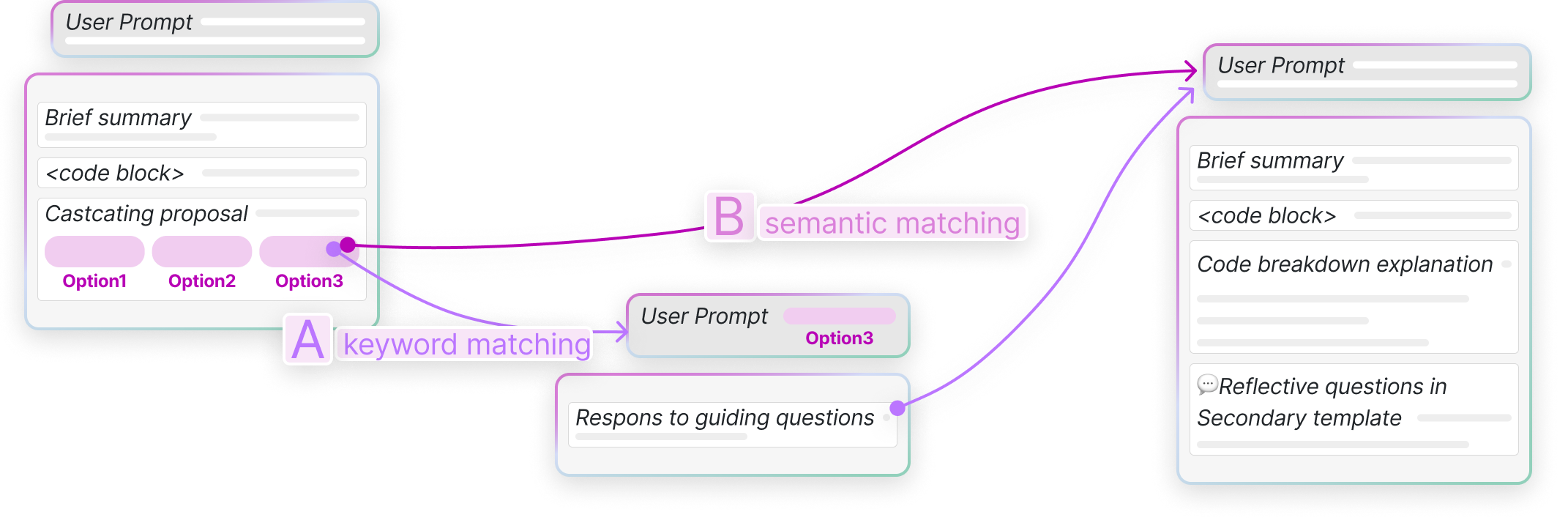}
    \caption{
        \hl{Secondary template supports two interactive ways: (1) prompts containing explicit option keywords trigger keyword matching (A); and (2) free-form prompts without explicit options invoke semantic matching (B).}
    }
    \label{fig:secondlevel-reflexacore}
\end{figure*}

Building on the reflection structures defined during the prototype development, \textit{Reflexa Core} as a dialogue panel (Figure~\ref{fig:teaser}-Core) enables three interaction modes (Figure~\mbox{\ref{fig:3Rs-template}}) with two-level reflective scaffolding (Figure~\mbox{\ref{fig:secondlevel-reflexacore}}). 
Rather than prescribing specific answers, each mode modulates the LLM’s response style to support different reflective needs during creative coding: 
\begin{itemize}
    \item \textbf{R1: Explainable \& Justified}, for articulating intent and reasoning
    \item \textbf{R2: Explorative (Multiple Perspectives)}, for divergent conceptual expansion
    \item \textbf{R3: Transformative}, for evaluating options or reframing goals from alternative perspectives
\end{itemize}
Each paired with templates tuned to the cognitive demand of the corresponding reflection type. We also design ``General'' mode to support dialogue without reflective questions.

\begin{figure*}[h]
    \centering
    \includegraphics[width=\linewidth]{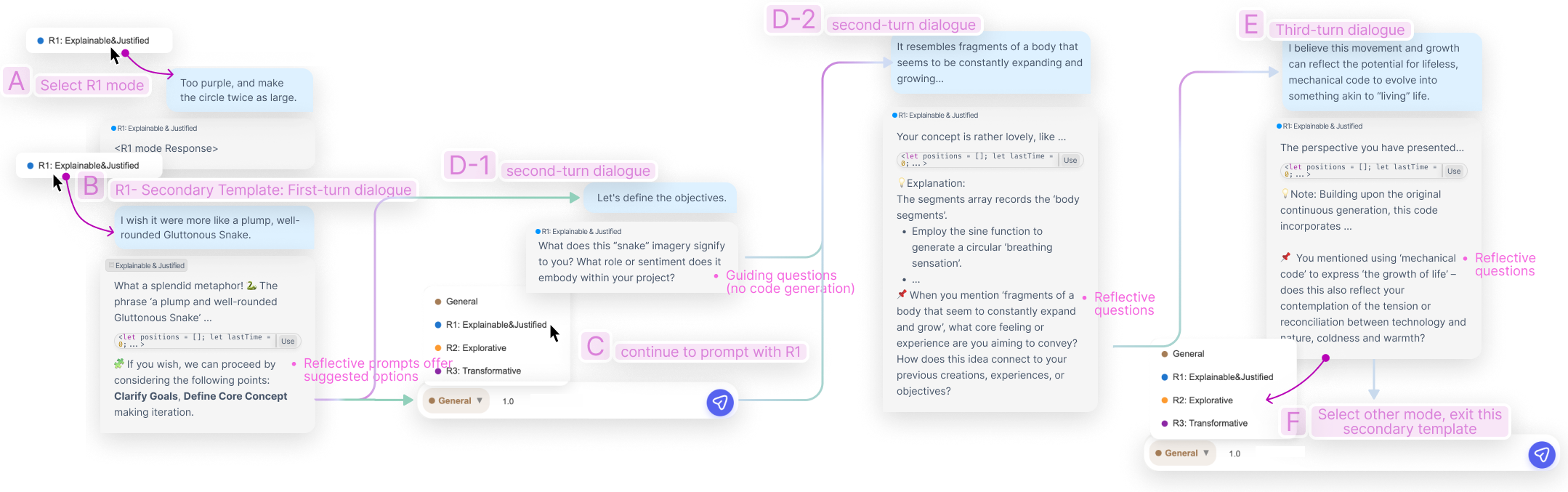}
    \caption{
        \hl{Interaction flow of \textit{Reflexa Core} when consecutive one mode (R1-R3) turns trigger the secondary template:} user prompts using R1 twice consecutively (A, B), and dialogue; (C) second-turn dialogue selecting Clarify Goals in two different ways (D-1, D-2); then response contains reflective questions to prompt further thinking (E), then exit the secondary template (F).
    }
    \label{fig:userflow-R1-second}
\end{figure*}

Under these three reflective modes, seven second-level LLM responses were designed tailored to dynamic and progressive process for encouraging deep reflection and further thinking (Table~\ref{tab:reflectionscaffold}): once one same reflection mode is used in a two-round dialogue, a curated structured template appears to guide further reflection. For instance, the second level scaffolding of R1 includes \texttt\small{Clarify Goal}-justifying creative motivation, and \texttt\small{Define Concept} (explaining goals behind functional code). 
The system automatically selects a template based on user inputs and then initiates either a follow-up LLM query or a complete response using the secondary template within the current reflection mode (as shown in Figure~\ref{fig:secondlevel-reflexacore}). Figure~\ref{fig:userflow-R1-second} demonstrates the interaction flow for a use case.
This design supports articulated intentions and conceptual breakdowns, aligning with DG2 (Scaffold Intent Articulation) and DG3 (Elevate Low-level Breakdowns). 

\begin{table*}[h]
\caption{\normalsize{
    \hl{
        Seven secondary templates designed in Reflexa Core, following R1 (Explanable\&justified), R2 (Explorative), and R3 (Transformative), using three strategies (a) (b) (c) in dialogue-based reflection scaffolds. 
    }
}}
\footnotesize
\noindent
\begin{minipage}[t]{0.49\textwidth}
\vspace{6mm} 
\raggedright
\begin{tabularx}{\textwidth}{>{\raggedright\arraybackslash}p{0.15\textwidth}X}
\multicolumn{2}{p{\textwidth}}{\normalsize{\textbf{(a) Externalize and structure intention articulation through reflective prompting }}} \\
\hline
\small\textbf{R1 | Clarify Goal} & 
    \textbf{Def.} Explore vague inspiration into concrete direction. \newline
    \textbf{Template} Help clarify users' goals step by step. \newline
    \emoji{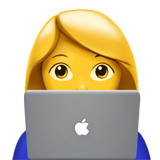}\textcolor{violet}{\textbf{Prompt:} \texttt{You are now a guided questioner, please help users clarify design goals in their creative project.}} \newline
    \emoji{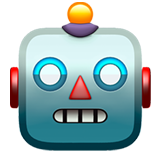}\textit{Exp Output:} \texttt{\uline{"Let's clarify your goals step by step:
    1) are you trying to express, document, or channel emotions? 
    2) what do you want the user to experience? 
    3) why is \color{brown}{'the subject'} important to you? 
    4) what form do you want the work to take?"}} \\
    
\midrule
\small\textbf{R1 | Define Core Concept} & 
    \textbf{Def.} Distill essential idea, mechanism, or metaphor \newline
    \textbf{Template} Help users explain what visual or interaction experience each key function is meant to create. \newline
    \emoji{female-technologist}\textcolor{violet}{\textbf{Prompt:} \texttt{Focus on the central concept and its relationship to the artist's creative drive.}} \newline
    \emoji{robot}\textit{Exp Output:} \texttt{
        \uline{1) "Does this \color{brown}{'sense of flow'} relate to a \color{violet}{particular emotion} or \color{violet}{user experience}?}\newline
        \uline{2) What is the idea of \color{brown}{'sense of flow'} for or from?"}} \\
        
\midrule
\small\textbf{R1 | Justify Detail Decisions} & 
    \textbf{Def.} Make users articulate the rationale behind detailed visual and interaction tweaks. \newline
    \textbf{Template} Help users explain why they changed specific animations, colors, timings, or micro-interactions. \newline
    \emoji{female-technologist}\textcolor{violet}{\textbf{Prompt:} \texttt{Help users justify and refine their detailed design decisions such as color, animation, rhythm, and layout.}} \newline
    \emoji{robot}\textit{Exp Output:} \texttt{\uline{"
        1) How does \color{brown}{'this detail'} reinforce the overall mood or message of \color{brown}{'the work'}? 
        2) Does \color{brown}{'this change'} make the experience clearer, richer, or more distinctive for the viewer? 
    "}} \\
\end{tabularx}

\vspace{0.3em}


\end{minipage}
\hfill
\begin{minipage}[t]{0.49\textwidth}
\vspace{0mm} 
\raggedright
\begin{tabularx}{\textwidth}{>{\raggedright\arraybackslash}p{0.16\textwidth}X}
\multicolumn{2}{p{\textwidth}}{\normalsize{\textbf{(b) Simulate future states and compare alternatives to scaffold anticipatory reasoning.}}} \\
\hline

\small\textbf{R2 | Conceptual Connections} & 
    \textbf{Def.} Expand ideas by linking inspirations and references. \newline
    \textbf{Template} Guide users to identify relationships, synergies, and hybrid concepts across inspirations. 
\newline
    \emoji{female-technologist}\textcolor{violet}{\textbf{Prompt:} \texttt{You are now a conceptual connector. Help users explore links between inspirations.}} \newline
    \emoji{robot}\textit{Output:} \texttt{\uline{"1) What themes exist in your inspirations? 2) Any tensions or synergies? 3) What new narratives emerge when combined? 4) How would you describe the hybrid?"}} \\ 
    \midrule
    
\small\textbf{R2 | Module–Experience Relations} & 
    \textbf{Def.} Connect code modules to narrative/experiential flow. \newline
    \textbf{Template} Guide users to examine how modules coordinate and influence experience. \newline
    \emoji{female-technologist}\textcolor{violet}{\textbf{Prompt:} \texttt{You are now a systems explainer. Help users relate modules to experience.}} \newline
    \emoji{robot}\textit{Output:} \texttt{\uline{"1) What are the key modules? 2) What experience does each support? 3) How can they coordinate better? 4) How would re-grouping change the experience?
"}} \\
\end{tabularx}

\vspace{1em}

\begin{tabularx}{\textwidth}{>{\raggedright\arraybackslash}p{0.16\textwidth}X}
\multicolumn{2}{p{\textwidth}}{\normalsize{\textbf{(c) Translate abstract ideas into visual-functional primitives via analogy and decomposition.}}} \\

\midrule
\small\textbf{R3 | Transform Creative Direction} & 
    \textbf{Def.} Reframe concept via new tone, angle, or audience. \newline
    \textbf{Template} Guide users to reimagine their direction from alternative emotional or narrative frames.  \newline
    \emoji{female-technologist}\textcolor{violet}{\textbf{Prompt:} \texttt{You are now a creative reframer. Help users shift perspective.}} \newline
    \emoji{robot}\textit{Output:} \texttt{\uline{"1) What changes if you adopt a different emotional tone? 2) How would visuals adapt? 3) What shifts with a different audience? 4) Which alternative direction feels promising?"}} \\
    
\midrule
\small\textbf{R3 | Re-evaluate and Shift Visual Style} & 
    \textbf{Def.} Reassess whether current style fits core intention. \newline
    \textbf{Template} Help users step back from local details to evaluate and, if necessary, reorient the overall style. \newline
    \emoji{female-technologist}\textcolor{violet}{\textbf{Prompt:} \texttt{You are now a style critic. Help users reevaluate visual style.}} \newline
    \emoji{robot}\textit{Output:} \texttt{\uline{"1) What core feeling should the style convey? 2) Where does it align or drift? 3) Which parts can change? 4) Describe a revised style in one sentence"}} \\
\end{tabularx}

\vspace{2.7em}

\end{minipage}
\label{tab:reflectionscaffold}
\end{table*} 

\subsubsection{Reflexa Flow: Non-linear Iteration and Reflective Traceability}
\begin{figure*}[h]
    \centering
    \includegraphics[width=\linewidth]{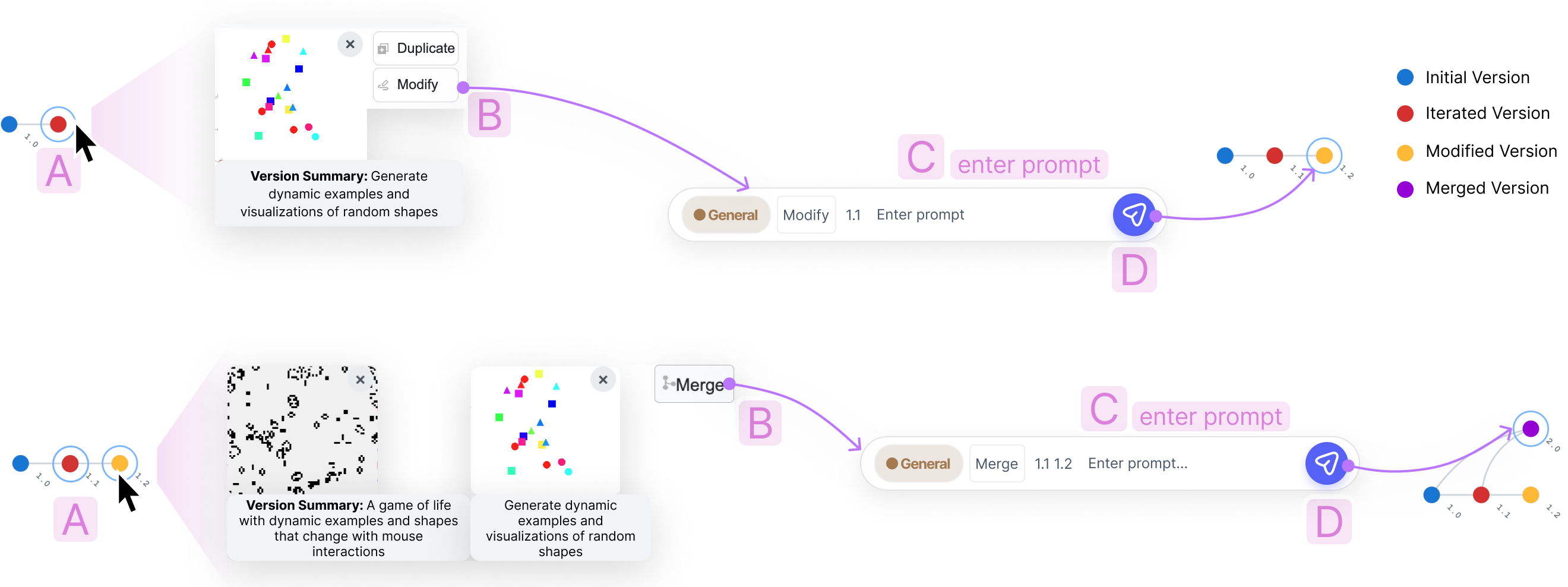}
    \caption{\hl{Interaction flows for the modify and merge operations in the \textit{Reflexa Flow}. Users (A) select one or two node(s), (B) choose modify or merge, (C) input a prompt, and (D) submit the operation.}
    }
    \label{fig:modify/merge}
\end{figure*}

\textit{Reflexa Flow} provides a visual node-based workspace (Figure~\ref{fig:teaser}-Flow) that supports non-linear exploratory creation through version saving, branching, and operations. Key features include: 
(1) \textbf{Version branch tree}: review and compare collected versions via previews by clicking a node, linking the corresponding dialogue history, visual output, and code;
(2) \textbf{Node-based operations}: delete, duplicate, modify, and merge versions to flexibly create multiple branches, facilitating exploration of diverse creative directions. Figure~\ref{fig:modify/merge} shows user interaction flows under modify and merge in the Reflexa Flow. 

This flow supports comparison-based reasoning, retrospective reflection, and iterative decision-making, addressing reflection gaps in trial-and-error cycles (C4) and aligning with DG1 (Support Non-linear Reflective Loops). 

\subsubsection{Reflection Sparks: Lightweight Triggers for Fast Reflection}~\label{sec:reflexa-sparks} 
\begin{figure*}[h]
    \centering
    \includegraphics[width=0.95\linewidth]{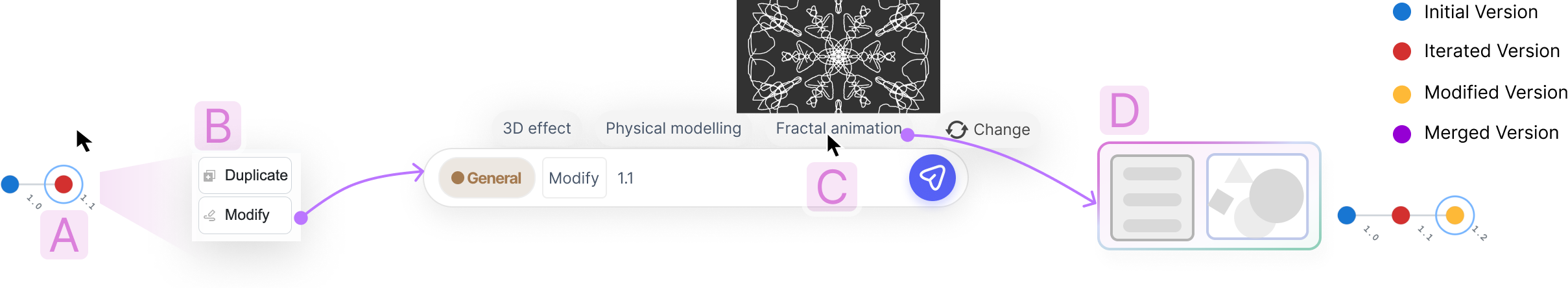}
    \caption{\hl{Interaction flow for \textit{Reflexa Spark} proceeds as follows: users (A) selects a node, (B) choose modify, and (C) activates the Spark options from the pop-up menu. Once user select one, the p5.js panel automatically (D) displays the corresponding code and preview, along with a saved version node in the Flow panel.}}
    \label{fig:Spark}
\end{figure*}
\textit{Reflection Sparks} embeds low-friction reflective buttons beneath the input area (Figure~\ref{fig:Spark}). After clicking ``modify'' on one selected version node in \textit{Reflexa Flow}, context-sensitive \textit{Spark} options suggest transformations, which were pre-defined and selected from \texttt{p5.js} reference~\footnote{\url{https://p5js.org/reference/}} (e.g., \texttt\small{3D effect}, \texttt\small{Fractal animation}), with visual previews. 
Clicking a Spark initiates iterative generation based on the selected reference, facilitating smooth transition from divergent ideation to deliberate refinement. 
This mechanism helps elicit structured responses from open-ended creative directions, particularly addressing early-stage ambiguity or hesitation (C1). 

\subsection{System Implementation}
The foundation of \textit{Reflexa} is a set of LLM-empowered API calls. They are built on the following technical pillars: memory management, predefined prompt templates, dynamic prompting strategies, context-aware version control, and AI-generated codes and reflections. The complete prompts for the \textit{Reflexa} are listed in the Appendix~\ref{apx:systemprompt}.

\subsubsection{Prompt design for LLM}
To support reflection and generate code by LLM, 
we design clearly structured prompts and apply serval prompting methods (e.g., few-shot prompting, Chain-of-Thought~\cite{cot} (COT) for \textit{Reflexa} system. 
The abovementioned four prompts (R1-R3, and General) is provided in the Appendix~\ref{apx:3Rprompts}).

These four main and seven second-level system prompt templates are built on the following elements:
 1) \textbf{Precise Persona Definition:} Each prompt begins by defining the LLM's core role as a p5.js expert and a creative guide for artists. 2) \textbf{Clear Theme Interpretation:} The prompts then outline a distinct task theme and core goal, guiding the LLM to focus on a particular mission. 3) \textbf{CoT Guidance} 
 that guides the LLM to complete the task by breaking down complex tasks into manageable stages. 4) \textbf{Few-Shot Prompting:} to guide the LLM to output content that aligns with the desired aesthetic and intellectual depth. 5) \textbf{Strict JSON Output:} States that a valid JSON object is expected, ensuring parsable and required fields in the output.

Beyond basic generation, the complex operations in \textit{Reflexa}—specifically ``merge'' and ``modify''—are designed to support reflection by exposing the underlying decision-making process (Appendix~\ref{apx:prompt-merge},~\ref{apx:prompt-modify}). 
Rather than functioning as a ``black box'' that simply outputs code, the system employs CoT prompts to mandate the LLM to articulate explicit rationales alongside the detailed implementation of the operations. 
For instance, in a ``merge'' operation, the LLM is guided to explain how it balances the visual grammar of two sources; similarly, for ``modify'', it clarifies how abstract intents are translated into code changes. 
By externalizing this reasoning, \textit{Reflexa} provides users with a clear and transparent logical trajectory of the code evolution during complex modifications. 

\subsubsection{Context management and version control}
In \textit{Reflexa} system, context management and version control are achieved through the collaboration of the frontend and backend. In the interface, \textit{Reflexa} uses a version graph to visualize and display chat history for only activating branches, providing a focused view of the conversation. And \textit{Reflexa} utilizes a ``Context Manager'' to track each version's data, including its code, chat logs, and explore the connections and evolve between versions. This provides a stable, traceable data flow for complex operations like ``merge'' and clear context when switching versions.

\subsubsection{System Implementation}
The \textit{Reflexa} creative coding system is implemented as a web-based application. The user interface is developed by the Vue 3~\footnote{\url{https://vuejs.org/}} framework in Javascript. To be exact, we leverage the Monaco Editor~\footnote{\url{https://microsoft.github.io/monaco-editor/}} as the interactive editor in the web application. The server side is constructed in Python, following a RESTful API architectural pattern. In particular, we use Langchain~\footnote{\url{https://www.langchain.com/}} to build an end-to-end pipeline of API-based LLM service from \textbf{gpt-4o} model and \textbf{text-embedding-ada-002} model provided by Mircrosoft Azure Cloud Service to vector database empowered by ChromaDB~\footnote{\url{https://docs.trychroma.com/}}. Moreover, we adapte 20 inspiration codes as example from official library of P5.js website.

\section{User Study}
    ~\label{sec:userstudy}
To examine \hl{the system-level impact of reflection scaffolding} in LLM-assisted creative coding, we conducted a within-subjects study \hl{comparing a holistically scaffolded reflection system—\textit{Reflexa}—with a baseline LLM tool intentionally designed to represent the common, unsupervised one-turn coding workflow without any reflective support.}

\subsection{Participants}

We recruited 18 artists with prior experience in creative coding and LLMs (Appendix~\ref{apx:participants}, Table~\ref{tab:participants}) via purposive sampling through social media (Xiaohongshu, Weibo) and generative art communities. 
Interested individuals completed an online screening questionnaire collecting demographics (age, gender, profession), computational art experience (duration, field), and GenAI tool usage. 
Inclusion required prior experience with \textit{Processing} or \textit{p5.js} and LLMs, ensuring familiarity with both technical and conceptual aspects of the study. Participants received \$14 USD for approximately 120 minutes. The protocol, including recruitment, consent, and data handling, was approved by the university Institutional Review Board (IRB).

\subsection{Study Design} 
\subsubsection{Methodological Positioning} 
\hl{This study is designed as a system-level evaluation comparing the presence versus absence of reflection scaffolding, rather than a component-wise ablation study. 
~
Reflection in creative coding is not enacted through isolated interface elements but through a constellation of interconnected processes—including moment-to-moment interpretation, hypothesis formation, exploratory variation, and goal reframing—identified in our formative study (Section~\mbox{\ref{sec:reflectionpractices}}). Therefore, \textit{Reflexa} instantiates reflection as an integrated construct, combining structured prompts (R1–R3), version-based navigation, and manipulation mechanisms into a unified reflective workflow. 
~
Testing individual components in isolation would produce ecologically invalid interactions by disrupting the reflective dynamics observed in real creative coding practice. Our methodological choice thus evaluates whether a holistically scaffolded environment meaningfully changes reflective behavior and creative outcomes relative to a standard LLM-based coding workflow without such scaffolding.
}

\subsubsection{Goal and Method}
We employed a within-subjects design in which each participant interacted with two systems: a baseline LLM-based creative-coding assistant without additional reflection scaffolds and the \textit{Reflexa} system integrated multiple reflection scaffolds). 

\subsubsection{Baseline}
\hl{
The baseline system functioned as a control condition representing the prevailing pattern of LLM-based creative coding: a single-threaded chat interface that generates code through one-shot or incremental prompts without structured reflection support. It retained the same LLM model, code editor, and preview environment as Reflexa to ensure equivalence in generative capability. To isolate the effect of reflection scaffolding, the baseline deliberately excluded all mechanisms that externalize or structure reflective reasoning—namely the tiered reflective prompt modes (R1–R3), the version-node visualization, and version manipulation functions (e.g., merge, modify), and direct visual outcome generation. 
~
These features are not general-purpose conveniences but constitute the reflective scaffold itself: version mapping enables comparison and explanation, manipulation supports hypothesis testing and strategy shifts, and structured prompts foreground goal articulation and reframing. Removing these components therefore produces a faithful representation of current LLM creative-coding practice while allowing the comparison to target the systemic presence of reflection scaffolds, rather than interface richness or feature quantity. A minimal “collection” button remained only to mark sketches for later self-assessment, without providing any reflective affordances. 
} 

\subsubsection{Study Design}

We employed a within-subjects design in which each participant used both the baseline system and \textit{Reflexa}. A within-subjects design is methodologically necessary for evaluating reflective scaffolds in creative coding. Creative style, risk-taking tendencies, and exploratory depth vary substantially across individuals; a between-subjects ablation design would therefore introduce more variance than insight. By having each participant engage with both systems, we control for these large individual differences, enabling a cleaner comparison of how the presence of reflection scaffolding shapes reflective behavior and creative outcomes. 
Each participant completed four exploratory creative-coding sketch variations per system, resulting in a total of eight sketches. The order of system use (i.e., which system was used first) was counterbalanced across participants to mitigate order effects.

\hl{
We adopted the \textit{Reflection in Creative Experience (RiCE)} questionnaire to measure reflection, as it captures moment-level reflective processes—interpretation, experimentation, and goal reframing—that also characterize creative coding. Because creative code evolves dynamically and often unpredictably, creators must continuously make sense of system outputs and adjust intentions in real time. RiCE therefore aligns with this micro-level reflective nature~\mbox{\cite{ford_reflection_2024}} and provides a valid measurement in our study context.
}

The independent variable was \hl{the presence of reflection scaffolding} (\textit{Reflexa} vs. baseline). Dependent variables included: 
(1) reflective behaviors elicited by each system; 
(2) evaluate participants' interaction experience and perceived agency when working with the systems; and 
(3) assess the creative outcomes produced across conditions (e.g., novelty, originality, aesthetic quality, complexity, and completeness). 

\begin{figure*}[h]
    \centering
    \includegraphics[width=1\linewidth]{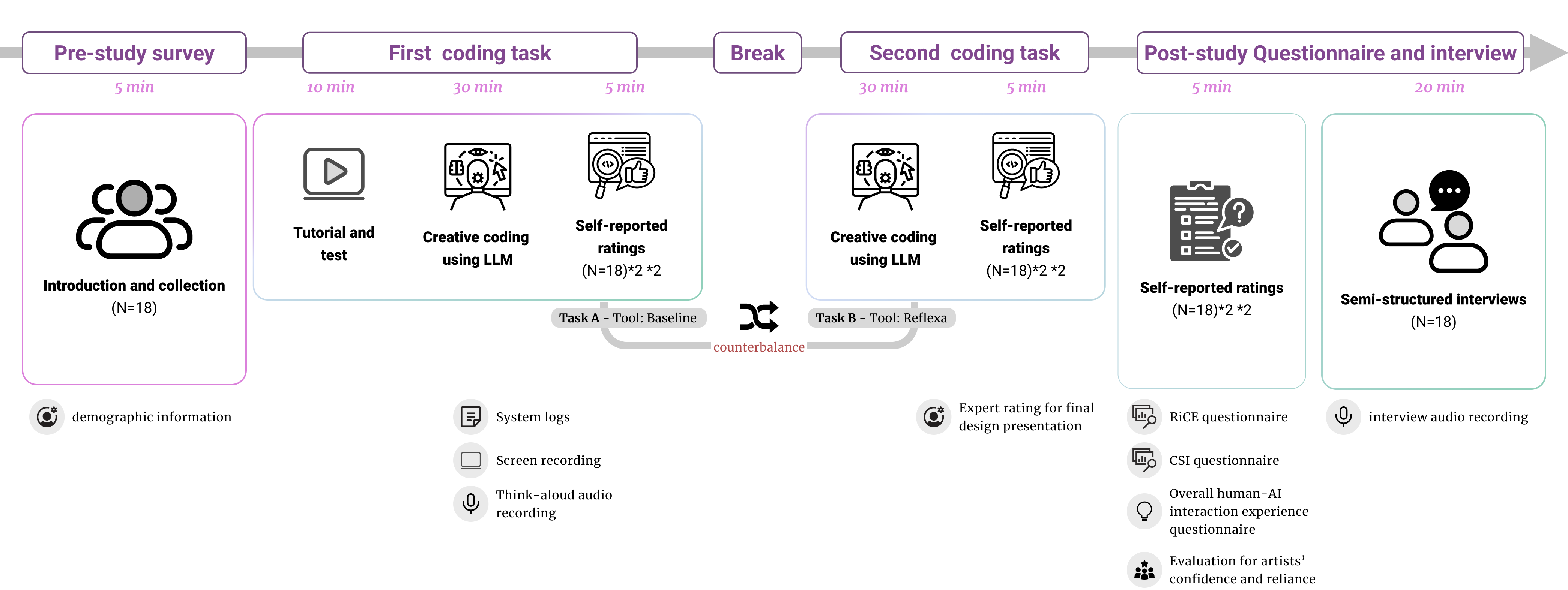}
    \caption{Study procedure of the creative coding experiment using a within-subjects design with 18 artists, comparing integrated reflection scaffolds in \textit{Reflexa} against the baseline system.}
    \label{fig:studyprocedure}
\end{figure*}

\subsection{Procedure}
Figure~\ref{fig:studyprocedure} summarizes the overall procedure. The session lasted approximately 120 minutes, including a 20-minute break, and comprised three stages: an introduction and tutorial, two timed creative-coding sessions (one per system), and post-task feedback including questionnaires and interviews.

    \subsubsection{Introduction (\textasciitilde10 min)} 
        Participants completed informed consent and received an overview of the study goals and procedure (see Figure~\ref{fig:studyprocedure}). They then viewed brief tutorial slides that demonstrated either \textit{Reflexa} or the baseline system (5 min) and were given 5 minutes to freely explore the assigned interface before beginning the coding session. \hl{Participants were instructed to engage in creative coding as they normally would in LLM-assisted workflows, with full access to the available features to preserve ecological validity and capture authentic reflective behaviors.}

    \subsubsection{Creative Coding Tasks (\textasciitilde60 min)} 
        Each participant completed two 30-minute creative-coding sessions, one with the baseline system and one with \textit{Reflexa}. During each session participants were asked to produce a working visual sketch in p5.js that: (1) had an open thematic prompt (creative freedom), (2) included at least one interactive element, and (3) exposed at least two adjustable parameters (e.g., color, motion, opacity). These constraints provided comparability while preserving creative freedom. Participants could iteratively refine their sketches during the 30-minute period. Condition order was counterbalanced across participants.

    
    \subsubsection{Post-Task Feedback (\textasciitilde25 min)} 
    After each session participants selected from their collected sketches and provided self-assessments for the artifacts they considered most representative. 
    Once both sessions were complete, participants completed a battery of questionnaires (described below, approx. 5 min) and engaged in a semi-structured interview (approx. 20 min). The interview probed: (1) interactions with \textit{Reflexa}'s features (Core dialogue, Spark, Flow); (2) whether prompts triggered new perspectives, goal reconsideration, or plan changes; (3) comparative usefulness of \textit{Reflexa} versus the baseline; (4) memorable or critical moments when \textit{Reflexa} influenced creative thinking; and (5) suggestions for future reflection-support features (full interview guide in Appendix~\ref{apx:interview-guide}).

    Following the interview, participants completed self-report questionnaires using 7-point Likert scales \cite{joshi2015likert}. Instruments included the \hl{Reflection in Creative Experience (RiCE) Questionnaire} measured three reflections, \Cp{} (Cp), \Se{} (Se), and \Ex{} (Ex) ~\cite{ford_towards_2023} (Appendix~\ref{apx:UserStudy-RiCE}, Table~\ref{tab:RiCE}), the Creativity Support Index (CSI)~\cite{cherry_quantifying_2014} (collaboration, enjoyment, exploration, expressiveness, immersion, result worth effort)), and a custom questionnaire assessing overall human–AI interaction experience (perceived controllability, transparency, cognitive load, collaboration, trust; Appendix~\ref{apx:overallhuman-AI}, Table~\ref{tab:HACquestions}). Participants also reported their confidence in, and reliance on, the AI system (Appendix ~\ref{apx:agency}). Finally, participants rated their creative outcomes on 7-point Likert scales for novelty, originality, complexity, aesthetics, completeness, interpretability, evolution, and satisfaction \cite{10.1145/1323688.1323689,Runco01012012,Carson01022005}.
    Finally, three expert evaluators independently assessed the creative coding outcomes using the same scale across novelty, originality, complexity, aesthetics, and completeness. Their evaluations were used to complement participant self-assessments and provide an objective benchmark for outcome quality.

\subsection{Data Analysis}
For quantitative data comparing \textit{Reflexa} and the baseline system within the same participants, we first assessed normality of difference scores using Shapiro–Wilk tests. Paired-sample $t$-tests were conducted when normality held; otherwise, Wilcoxon signed-rank tests were applied. This approach evaluated overall system effects on creative outcomes, interaction experience, and agency. 


To investigate how reflection impacted creative outcomes, we first analyzed both self- and expert-rated metrics. Expert ratings were validated with high inter-rater reliability (ICC: \textit{Novelty} = .946, \textit{Originality} = .975, \textit{Completeness} = .912, \textit{Aesthetic} = .938, \textit{Evolution} = .922, \textit{Complexity} = .916; all $p < .001$), reflecting quantitative consistency across raters and ensuring the quality of evaluation. 
Self- and expert-rated scores were averaged across the four successive task versions to capture overall process outcomes, and iterative improvements were analyzed using non-parametric Friedman and Page tests, suitable for ordinal self-reports and small sample size ($N = 18$). 
We also conducted system-specific correlation analyses among creative outcomes, reflection scores, as well as interaction and agency, using Pearson correlations when normality assumptions were met and Spearman correlations otherwise.

Qualitative interview, prompt, and usage recording data were thematically analyzed~\cite{Braun01012006} using a combined inductive-deductive approach~\cite{jennifer2006thematic}, integrating participants’ lived experiences with our research objectives. 
Data were segmented by interview questions, prompt-response pairs, and recorded coding actions to define consistent coding units. 
Two researchers independently conducted open coding, identifying initial codes from participants’ statements and system interactions. Discrepancies were resolved through discussion, resulting in a collaboratively refined codebook with hierarchical themes. Deductive coding categories were informed by established reflection types (\textit{Cp, Se, Ex}), while inductive codes emerged from recurring patterns in participants’ descriptions and behaviors. Each category was further aligned with quantitative metrics (e.g., RiCE, CSI, outcome measures) to integrate mixed-method insights. Iterative discussions and refinements finalized the codebook, which was then applied to the full dataset, producing main themes and subthemes capturing user benefits, human-AI collaboration experiences, and creative outcomes. Coding was conducted using Nvivo, ensuring systematic traceability and replicability.

\section{Findings}

\hl{Our evaluation adopts a mechanistic lens: 
we examine how Reflexa’s integrated scaffolds elicit specific forms of reflective behavior (Section~\mbox{\ref{sec:finding-reflection}}) and how these behaviors in turn relate to creative process and outcomes (Section~\mbox{\ref{sec:reflection->process}} and ~\mbox{\ref{sec:reflection->outcome}}). 
} 

\begin{figure*}[t!]
    \centering
        \begin{subfigure}[t]{0.32\textwidth}
        \centering
        \includegraphics[height=5cm]{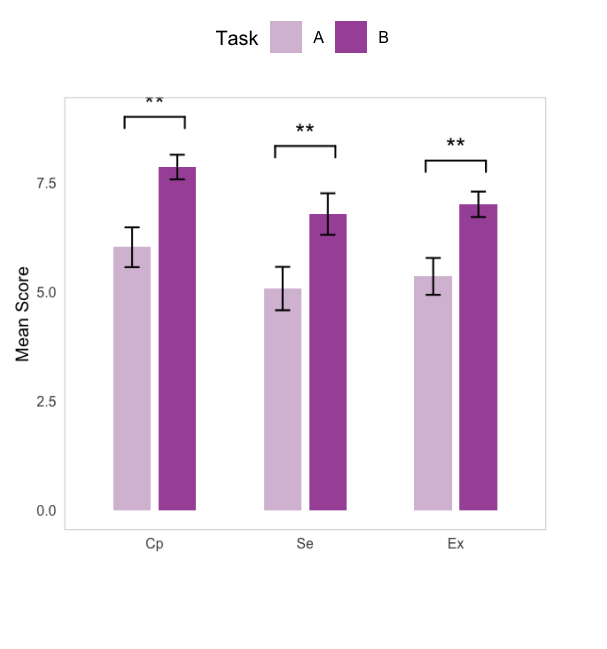}
        \caption{}
        \label{fig:RiCEbarchart}
    \end{subfigure}%
    ~
    \begin{subfigure}[t]{0.4\textwidth}
        \centering
        \includegraphics[height=5cm]{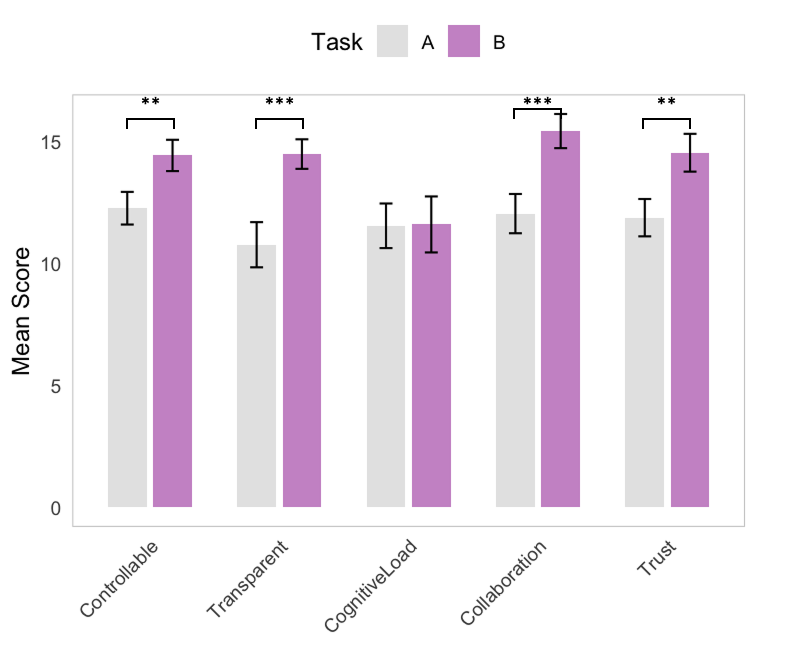}
        \caption{}
        \label{fig:HAC_comparison}
    \end{subfigure}%
    ~ 
    \begin{subfigure}[t]{0.27\textwidth}
        \centering
        \includegraphics[height=5cm]{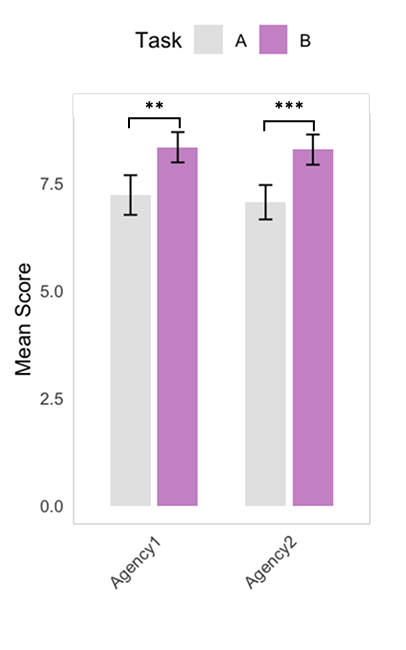}
        \caption{}
        \label{fig:agency_comparison}
    \end{subfigure}
    \caption{\hl{Participants’ reflection, interaction experience, and agency when using \textit{Reflexa} (Task B) versus the baseline (Task A).
(a) Reflection levels across Cp, Se, and Ex modes; (b) overall AI interaction experience in controllability, transparency, cognitive load, collaboration, and trust; and (c) artists’ agency in self-confidence (Agency1) and self-directed AI reliance (Agency2).}
}
    \label{fig:A2+agency-barchart}
\end{figure*}

\subsection{Reflexa-Elicited Reflective Behavior Mechanisms}~\label{sec:finding-reflection}

\hl{Our analysis reveals that \textit{Reflexa} elicited \emph{distinct reflective mechanisms} that were largely absent in conventional LLM-assisted coding.}
~
RiCE results (Figure~\ref{fig:RiCEbarchart}; Appendix~\ref{apx:findings}, Table~\ref{tab:RiCE_comparison})
indicated that Reflexa gained high scores across all 
dimensions-\textit{Cp}, \textit{Se}, and \textit{Ex}. 
Across 18 participants, \textit{Reflexa} elicited 252 prompts ($M_{Reflexa} = 14.00$, $SD_{Reflexa} = 3.40$), exceeding the baseline's 187 prompts ($M_{baseline} = 10.40$, $SD_{baseline} = 4.17$), with a mean difference of 3.60 prompts ($t(17) = 3.126$, $p = 0.006$**), indicating increased user engagement. Prompt length did not differ significantly between \textit{Reflexa} ($M_{Reflexa} = 17.90$, $SD_{Reflexa} = 18.80$) and baseline ($M_{baseline} = 15.80$, $SD_{basline} = 10.80$; $V = 93$, $p =0.766$). Qualitative results reveals Reflexa’s multi-layered scaffolds on reflection: structured and adaptive R1–R3 dialogues fostered articulation of intentions and iterative reasoning; visual version exploration supported mechanism tracing and hypothesis testing; and lightweight Spark encouraged fast experimentation during diverging creative decisions. 

\subsubsection{Insights from Reflection Strategies}
First, \textbf{dialogic reflection scaffolding} (\RC{}) through progressive multi-mode dialogues (R1–R3) enabled real-time feedback and guiding iterative idea refinement. 
Participants noted that the system dialogue \textit{``persistently probed with `why' and `what next', compelling me to articulate my thoughts clearly''} (P17), and remarked that such guidance \textit{``made me realise there were alternative directions I hadn't considered,''} (P13) illustrating that reflective depth emerged from dialogic framing rather than explicit instruction. 
Participants demonstrated: R1 supported stepwise articulation of goals and rationale, while R2 and R3 introduced perspective shifts and alternative framings that broadened conceptual space. As P9 observed: \textit{``Its issues make me think of entirely different approaches to performance.''} Together, these modes created a rhythm from explanation, reconsideration, to reformulation during moment-to-moment coding. 

Second, \textbf{visually node-based version exploration} (\RL{}) enabled participants to anchor reasoning in observable differences across iterations. 
Several participants described the role of version backtracking in reasoning. For instance, P13 noted: \textit{``I would revisit previous nodes, consolidate different attempts, and then decide on the next version.''}
Across \textit{Reflexa} sessions, participants typically produced multi-branch version trees with several levels of depth, frequently using duplication to fork alternative ideas while preserving prior states. Merge (N=8) and modify (variation-generating) (N=7) operations were concentrated in phases where participants explicitly described \textit{``trying a different direction''} or 
\textit{``seeing what happens if I combine these effects,''} aligning with experimental reflection. Each modified instance (N=8/8)—and 6/7 merged instances were further developed. 
Version-node analyses also revealed that participants performed this reasoning through distinct branching strategies (Figure~\ref{fig:branchingpatterns}). 
 Some advanced ideas through diversified linear extension with occasional revisits (e.g., P5, P8). Others pursued divergent branching to interrogate alternative mechanisms (e.g., P3, P7, P12, P13). Several practiced diversified iteration by merging outputs into refined solutions (e.g., P4, P11). These strategies demonstrate how version management became a cognitive scaffold for reasoning—allowing participants to anchor hypotheses in observable comparisons and orchestrate exploration with deliberate structure.  

Third, \textbf{rapid visual iteration} (\RS{}) facilitated efficient exploration of design alternatives. Participants noted that this process supported both exploration and fine-tuning of creative outcomes. For example, the feature was described as \textit{``something I could never imagine''} yet yielded \textit{``a weird but inspiring result that led me to iterate further''} (P11). 
Collectively, these practices enhanced cognitive engagement, and promoted active experimentation. 

\subsubsection{Behavior mechanisms elicited by Reflection} Qualitative analysis revealed several recurring mechanisms: 


\begin{figure*}
    \centering
    \includegraphics[width=0.99\linewidth]{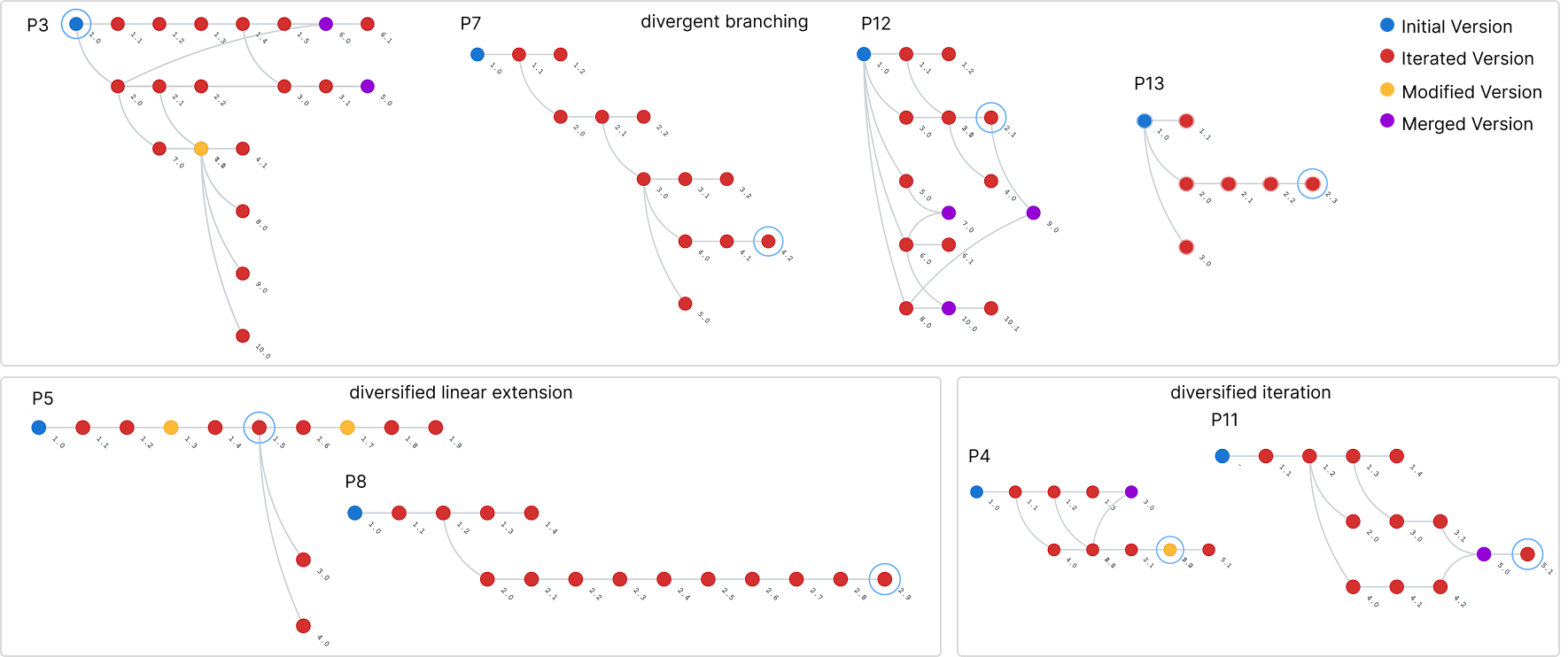}
    \caption{\hl{Selected version-node visualizations created in Reflexa Flow panel}, illustrating varied creative development strategies: diversified linear extension (P5, P8), divergent branching (P3, P7, P12, P13), and diversified iteration through merging and refinement (P4, P11).}
    \label{fig:branchingpatterns}
\end{figure*}

\paragraph{\hl{B1. Reflexa fostered generative-mechanism understanding (Cp, Ex).}} Participants increasingly interrogated why visual behaviors emerged, shifting from surface-level parameter adjustments to causal reasoning about the generative process. They articulated functional dependencies, diagnosed mismatches between intention and outcome, and developed explanatory accounts of system behavior. As P12 observed, \textit{``even a tiny parameter change completely shifted the outcome,''} reflecting heightened sensitivity to underlying mechanisms. This understanding was strengthened through iterative comparisons in \textit{Flow} and clarification prompts in \textit{Core}, enabling users to link parameter changes to perceptual effects. 
Reflexa thus supported a transition from intuitive fine-tuning toward a more systematic approach to constructing and validating structured hypotheses. 

\paragraph{\hl{B2. Reflexa fostered structured reflective reasoning (Cp, Ex).}}
Reflexa supported a shift from intuitive adjustments to systematic hypothesis formation and evaluation. Many participants proactively formulated hypotheses, test alternative approaches, and returned to version nodes to recover divergent pathways when encountering failures. As P13 noted, \textit{``by observing the patterns in different attempts, I can discern where it would be most reasonable to proceed next.''} 
The coexistence of parallel versions helped users discern structural differences and trace divergent paths, transforming reflection from retrospective summarization into real-time strategic adjustments.
This structured reasoning functioned as a cognitive scaffold for navigating uncertainty, coordinating dialogues, version trees, and experimental variants into coherent reflective cycles. 

\paragraph{\hl{B3. Reflexa fostered conceptual abstraction and goal reframing (Se, Ex).}} 
Beyond immediate visual adjustments, participants demonstrated higher-level reframing of creative direction—revisiting thematic intent, stylistic coherence, or experiential goals. Dialogic prompts in R3 prompted users to step back from local fixes and reconsider the conceptual grounding of their work. As P11 described, the process \textit{``made me rethink what I was actually trying to express.''} This abstraction enabled creative pivots, such as adopting new emotional tones, redefining narrative aims, or synthesizing hybrid stylistic elements, often triggered by unexpected merge outcomes. Such reframing reflects Se- and Ex- level reflection, extending reasoning beyond procedural adjustments toward purposeful reorientation of creative intent. 


Overall, these mechanisms illustrate how Reflexa scaffolded reflection not as isolated prompts but as a \emph{progressive, mechanism-oriented process} that deepened users’ understanding of their work and expanded the creative space. 

\subsection{How Reflective Behavior Mechanisms Reshape Creative Coding Workflows?}~\label{sec:reflection->process}

\begin{figure*}[h]
    \centering
    \includegraphics[width=\linewidth]{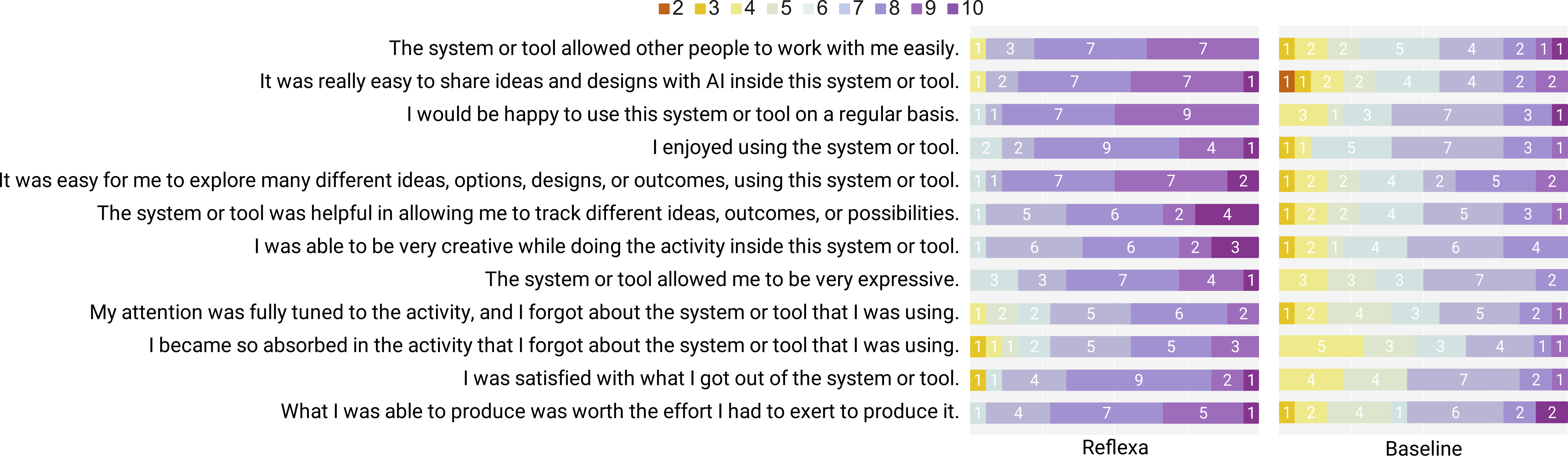}
    \caption{Likert scale of creativity support index (CSI) questions between \textit{Reflexa} and the baseline systems.}
    \label{fig:CSI-likert}
\end{figure*}

\hl{Building on the three behavior mechanisms identified in Section~\mbox{\ref{sec:finding-reflection}}, we examined how these reflective behaviors shaped interaction experience and creative supports. }
~\label{sec:reflection->process-interaction}
~\label{sec:reflection->process-agency}
\hl{Participants reported Reflexa provided creativity support, scoring CSI at 94.7 (SD = 10.5) — a high score for creativity support index~\mbox{\cite{cherry_quantifying_2014}}. As shown in Figure~\mbox{\ref{fig:CSI-likert} and Appendix~\ref{apx:findings}}, Table~\mbox{\ref{tab:CSI_comparison}}, while Reflexa scored relatively high in all dimensions, participants found it provided the strongest support for the exploration of ideas (M = 16.6, SD = 1.85). This trend emerged from reflection-enabled shifts in how participants interpreted, navigated, and extended their creative space.} 
Participants also reported high interaction quality across controllability, collaboration, transparency, and trust (Figure~\ref{fig:HAC_comparison}; Appendix~\ref{apx:findings}, Table~\ref{tab:HAC_comparison}), as well as high self-directed AI reliance (Figure~\ref{fig:agency_comparison}; Appendix~\ref{apx:findings}, Table~\ref{tab:agency_comparison}) during creative coding process.
~
Qualitative analyses revealed that the underlying causes stemmed from changes in AI interaction experiences and perceived agency triggered by reflective mechanisms.

~~~~~~~~~~~~~~~~~~~
~~~~~~~~~~~~~~~~~~~
\paragraph{\hl{Mechanistic navigation enabled deeper exploration}}
Mechanistic navigation helped participants plan their next steps in the generative space. Participants employed mechanistic understanding (B1) and structured reasoning (B2) in multiple ways: several participants used it to interpret why visual behaviors emerged, while others relied on it to diagnose mismatches between intention and outcome. P9 and P13 used this understanding to anticipate how parameter adjustments would propagate through the system; P12 and P15 drew on these cues to refine their exploratory strategies. As P9 expressed, \textit{``Each version tells me why it turned out that way.''} P13 similarly noted, \textit{``I began to anticipate why the system generated certain outputs, then collaborated with it''}
In contrast, without such mechanistic grounding, participants previously described exploration as reactive and uncertain. As predictability increased, participants gained stronger control over the generative process and reconceptualized the AI from an execution tool to a partner offering behavioral insight. 


\paragraph{\hl{Reflective alignment supported stable intent and direction-finding}}
Reflective alignment improved participants’ ability to maintain a coherent direction. Participants reported that their iterative strategies became clearer and more deliberate as they engaged in structured reflective reasoning (B2) and conceptual reconstruction (B3). This stability, in turn, enhanced their sense of controllability: rather than being guided by emergent outputs, they were able to steer AI suggestions toward their intended goals. 
This increased stability can be attributed to several factors.  
First, reflective reasoning prompted participants to repeatedly articulate and refine their evolving intentions. 
Second, clearer intent enabled participants to evaluate AI-generated suggestions more selectively. 

\paragraph{\hl{Articulating logic made interaction more transparent and predictable}}
Articulating generative logic significantly reduced the uncertainty participants experienced during interaction. 
This practice enabled participants to interpret why the system behaved as it did, and  encouraged them to form hypotheses about causal relationships, by employing generative-mechanism understanding (B1) and structured reflective reasoning (B2). 
    For instance, participants noted articulating hypothesis such as \textit{“Could this be caused by a shading override?”} helped participants anticipate system responses. 
Participants also reported fewer instances of “blind trial-and-error,” indicating that such articulation lowered the cognitive demand of interpreting unexpected outputs. 
This reduction in uncertainty can be attributed to several factors. 
First, articulating hypotheses prompted participants to externalize generative assumptions, which enabled them to track causal dependencies more reliably. 
Second, the AI’s clarifying responses—now grounded in these articulated hypotheses—provided more targeted feedback. 
Lastly, participants required fewer corrective cycles because articulated logic allowed them to issue more precise behavioral modification requests, streamlining the overall iteration process. 

\subsection{How Behavior Mechanisms Translate into Creative Outcomes?}~\label{sec:reflection->outcome}

\hl{This section examines how reflective behaviors shaped creative outcomes.}
Reflexa yielded high self-assessed and expert-rated creative outcomes (Appendix~\ref{apx:findings-comparision}, Figure~\ref{fig:deltaplot-B-A}, and Tables~\ref{tab:self+expert-rate_comparison}). Across both Reflexa and the baseline, self-assessment and expert ratings showed statistically significant upward trends across the four-version development sequence (Appendix~\ref{apx:findings-comparision}, Table~\ref{tab:trend-self} and ~\ref{tab:trend-expert}). 
~
\hl{Section~\mbox{\ref{sec:reflection->process}} revealed two recurring elements in observation and reports of the creative process: interaction experience and AI reliance. Our further qualitative analysis identified that these factors also contributes to creative outcomes.} 
\label{sec:CSI}

~\label{sec:outcomes} 
~\label{sec:self/expert-outcome}
~\label{sec:mediator-outcome}

\subsubsection{\hl{Interaction Experience Mediating Creative Outcome}}
\begin{figure*}[h]
    \centering
    ~
    \begin{subfigure}[t]{0.4\textwidth}
        \centering
        \includegraphics[height=1.7cm]{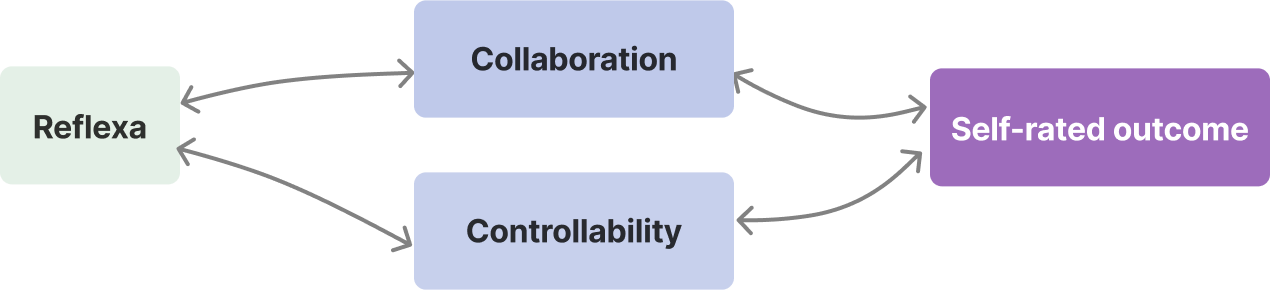}
        \caption{}
        \label{fig:mediator-self2}
    \end{subfigure}%
    \caption{Reflection mediating (a) controllability, and (b) collaboration in relation to self-rated creative outcomes.}
    \label{fig:mediator-self}
\end{figure*}

Our analysis indicates that controllability and collaboration not only shape participants’ creative processes but also directly influence their self-assessments of creative quality (Figure~\ref{fig:mediator-self}). These factors act as key mediators between reflection and creative outcomes by altering perceptions of “who is creating” and “how creation unfolds.”

\paragraph{\hl{Controllability grounds creative judgment in directional evaluation}}
\hl{Through reflective scaffolding, participants developed the ability to predict and regulate generative behavior, and this sense of mastery became the basis for directional evaluation of creative quality.} 
As participants gained confidence in predicting and adjusting the generative process, their evaluations increasingly centered on \textit{``whether I successfully achieved my intended expression''}.
This mastery encouraged more directional self-evaluation: participants emphasized \textit{``whether I maintained the coherence of the overall concept''} rather than assessing creativity solely on aesthetic grounds. Controllability therefore stabilized how participants attributed creativity, elevating their subjective assessments of work quality.

\paragraph{\hl{Collaboration fosters recognition of creative outputs as co-creative artefacts.}}
\hl{Reflection also transformed interaction into a form of co-reasoning, and this collaborative stance led participants to attribute greater intentional depth and coherence to the resulting artefacts.} 
Participants frequently described their collaboration with the AI as occupying a \textit{``shared cognitive space,''} which substantially elevated perceived creative value. 
Collaboration made reasoning processes transparent, enabling participants to discern \textit{``how ideas are formed.''} When participants viewed themselves as co-constructing ideas with the AI rather than merely responding to its outputs, they attributed greater \textit{``intentional depth''} to the work. 
Collaboration also enhanced a piece’s \textit{``coherence''} and \textit{``defensibility''}, as negotiating, supplementing, or rejecting AI suggestions strengthened justification for final decisions.
Thus, collaboration afforded participants interpretive agency over creative pathways, thereby elevating subjective evaluations.

\subsubsection{\hl{Agency Mediating Creative Outcome}}
\begin{figure*}[h]
    \centering
    \begin{subfigure}[t]{0.5\textwidth}
        \centering
        \includegraphics[height=1.2cm]{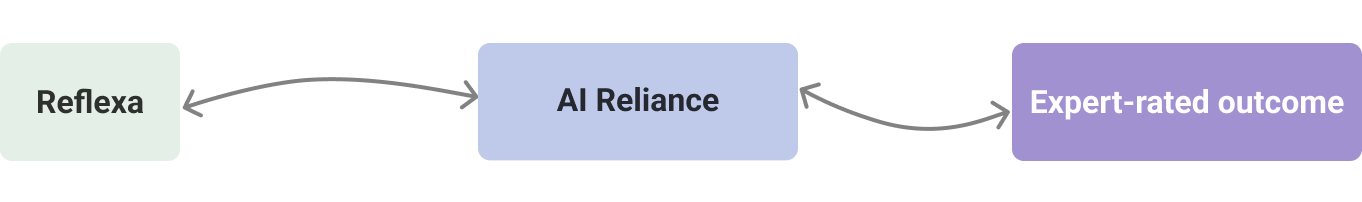}
        \caption{}
        \label{fig:mediator-e1}
    \end{subfigure}%
    ~
    \caption{Reflection mediating AI reliance in relation to expert-rated creative outcomes.}
    \label{fig:mediator-agency}
\end{figure*}

Beyond interaction experience, our results indicate the AI reliance also mediates the relationship between reflection and expert-rated creative quality (Figure~\mbox{\ref{fig:mediator-agency}}). 
\hl{Reflection shaped how participants balanced their own initiative with AI assistance, and this balance significantly influenced the structural sophistication of their creative outcomes.} 

\begin{figure*}[h]
    \centering
    \includegraphics[width=\linewidth]{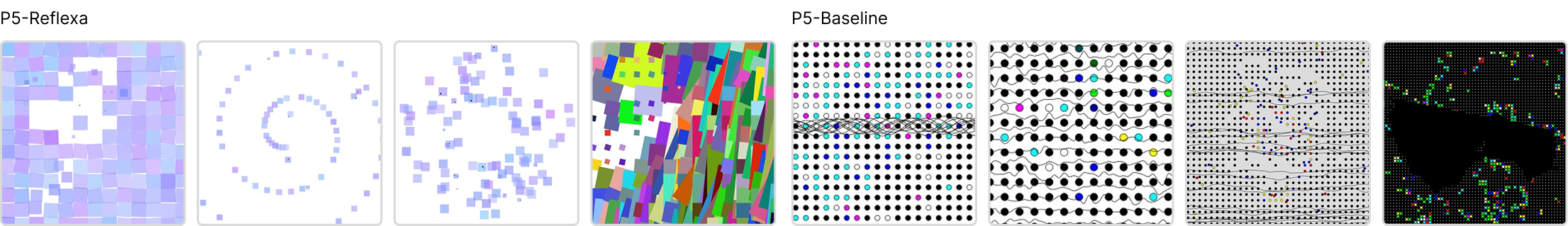}
    \caption{The variations of creative codes created by P5 in \textit{Reflexa} and baseline, explored different creative directions respectively.}
    \label{fig:p5-outcomes}
\end{figure*}

\paragraph{\textit{\hl{Better AI reliance enabled active exploration across multiple sources.}}}
\hl{Reflection encouraged participants to regulate their reliance on AI, and this self-directed stance enabled richer exploration across multiple sources.} 
Participants who did not depend solely on AI blended ideas, remixed outputs with their own concepts, and produced structurally more complex results. 
P5 exemplified this pattern (Figure~\mbox{\ref{fig:p5-outcomes}}): 
    Instead of accepting single-shot AI outputs, P5 repeatedly combined multiple sources of inspiration, layering new constraints and behaviors across prompts. Beginning with a simple gradient-color square, P5 progressively added new generative rules—\textit{``Color blocks randomly exhibit their own motion... rotating, bouncing, shifting within small ranges,''} followed by escape-like vanishing, ricocheting, scaling, and recomposition into rotational trajectories.
This sequence illustrates human-led structural evolution that reconstructed the generative architecture rather than asking the AI to elaborate on a single idea. 
In contrast, participants who relied heavily on AI were often limited to prompt tweaking or minor variations, struggling to achieve substantive structural changes.

\paragraph{\textit{\hl{Better AI reliance enabled effective steering of the AI.}}}
\hl{Reflection strengthened participants’ understanding of the model’s generative logic, enabling more effective steering toward desired outcomes.} 
Participants who developed such understanding adjusted constraints and behaviors with precision, producing more intentional and differentiated generative structures. 
    P5’s prompts exemplify this competence: after observing repetitive block animations, P5 explicitly regulated the generative logic through conditional structures—\textit{``Before the mouse cursor passes over a color block... it maintains its original motion state; when the cursor passes over a block... it vanishes in an evasive manner... subsequently returning to a random position.''} This demonstrates an understanding not only of what the AI generated but also why it behaved in that manner.  

\paragraph{\hl{Integrated pathway model}}
Together, these mediators outline a coherent pathway: Reflection strategies → reflection behavior mechanisms (B1–B3) → experiential and agency mediators → enhanced creative outcomes. 
This mechanism-linked pathway accounts for both self-rated and expert-rated creativity and explains why Reflexa outperformed the baseline: it more reliably activated reflection-driven processes that are inherently conducive to creative advancement.

\section{Discussion}

\hl{Compare to prior studies, our work is the first to empirically investigate how LLM-supported reflection operates and influences human creativity. 
Our findings reveal how \textit{Reflexa}—through elicited reflective behaviors—frames reflection as \textbf{a co-constructed cognitive process between human and AI, extending Schön’s human-only \textit{reflection-in/on-action}} (\mbox{\cite{schon1983reflective,schon2017reflective}}) toward the context of AI-mediated reflective systems.}

\subsection{Reflection as a Foundational Mechanism for Human–AI Co-reflection and Co-creation}

\subsubsection{Coupling Reflection and Cognitive Processes Toward Co-reflection.}
\hl{Our findings indicate that creative processes were shaped by \textbf{a mechanism-oriented form of reflection supported by deliberate, well-designed LLM scaffolding.} 
Whereas prior work (e.g.,~\mbox{\cite{ford_reflection_2024,wang2025pinning,xu_productive_2025}}) primarily documents “improvisatory” and “instinctual” reflective experience  grounded in self-descriptive narrative. \textit{Reflexa} demonstrates that reflection can extend into the \emph{computational substrate} of both AI and creative work. Its scaffolds surface generative logic, parameter–effect relationships, and causal dependencies that typically remain opaque, enabling participants to reason explicitly about the mechanisms driving algorithmic behavior. 
By making these, the scaffolding shifts reflection from narrative descriptions of what happened (i.e., Fleck and Fitzpatrick's reflection theory~\mbox{\cite{fleck_reflecting_2010}}) to explanations of why and how outcomes happened. 
}


\hl{This shift aligns with dual-process of creativity, which emphasize the interplay between intuitive exploration and deliberate, rule-based reasoning\mbox{~\cite{SowdenPaul2014TSSo,smith2000dual}}. 
Reflexa’s scaffolds \textbf{externalize the rational components} of this process—such as hypothesis formulation, evaluation of causal dependencies, and interpretation of system responses—that are usually implicit, fragmented, or effortful to maintain.  
This externalization creates cognitive capacity for intuitive experimentation while preserving awareness of goals and constraints, fostering fluid between deliberate analysis and exploratory improvisation. 
More broadly, these dynamics suggest that reflection in human–AI co-creative systems operates as a form of distributed metacognition—mediating how participants interpret system behavior, negotiate design intentions, and adjust strategies over time. 
}

\subsubsection{Reflection as a Coordination Mechanism for Human–AI Co-creativity.}
Our findings indicate that reflection reshaped how participants interpreted, steered, and evaluated the creative process, which in turn impacts creative outcomes.
\textit{Reflexa} functioned less as a prompt generator and more \textbf{as an interactive partner through which user negotiated controllability, collaboration, and agency (Section~\mbox{\ref{sec:mediator-outcome}}).}
%

\hl{This perspective advances prior accounts of human-AI creative collaboration, which have largely focused on agency distribution or assistance (e.g.,\mbox{~\cite{xu_productive_2025,shen_ideationweb_2025}}). 
Our study highlights reflection scaffolding as a coordination mechanism that structures how users and the model jointly navigate evolving creative intentions. 
In Reflexa, users engaged in collaborative sensemaking—articulating rationales, interpreting generative behaviors, and aligning subsequent steps with the system’s suggestions. 
This highlights a complementary layer of collaboration: not only \emph{who} acts, but \emph{how} joint reasoning unfolds. 
Reflective structuring stabilized exploration, clarified intent, and enabled more deliberate shared control. Consequently, our work expands the design space of human–AI creative collaboration by framing reflection as a critical organizational resource for coordinating intentions, mechanisms, and actions during co-creation. 
} 

\subsection{Temporal Constraints and the Depth of Reflection in AI-Supported Creative Coding}
\hl{While Reflexa supports moment-to-moment reflection-in/on-action during creative coding, deeper forms of creative reflection are intrinsically temporal and unfold over extended periods, as observed in many other creative domains. 
In practice-driven tasks, participants often prioritized visible progress through intuition and rapid iteration, which constrained opportunities for meta-cognitive engagement, such as reflection on self. Investigating longer-term reflective processes may therefore broaden reflection theory and extend our understanding of co-reflective dynamics with AI.

To address these temporal constraints, frameworks such as longitudinal reflection~\mbox{\cite{Kjærup2021}} and experience-centered design~\mbox{\cite{wright2010experience,10.1145/1460355.1460360}} offer approaches for sustaining reflection over time. Mechanisms including reflection diaries and AI memory cues~\mbox{\cite{wu2025humanmemoryaimemory,10973813,10892993,10.1145/3757616}} can connect momentary insights to longer reflective trajectories, deepening meta-cognitive awareness and fostering ongoing human–AI co-reflection. 
Future work should explore adaptive prompting and reflective-trajectory visualizations to integrate immediate creative flow with sustained reflective growth and shared agency.}

\subsection{Design Implications} 
\subsubsection{\textit{Systems should function as reflection–creativity couplers rather than passive scaffolds.}} 
\hl{First, given we found the linking between different reflection and interaction experience, and agency in human-AI co-creation, this offers a critical lens on collaborative ways and agency, contributes to other HCI studies on human-AI collaboration.} 
These results suggest \textit{Reflexa} act as couplers rather than neutral scaffolds, actively shaping interactions between cognitive processes and generative outcomes \cite{clark2008supersizing, hollan2000distributed}. Future systems could incorporate domain-specific adaptations and sustained support to maintain effective human–AI cognitive alignment across extended creative sessions.

\subsubsection{\textit{System should guide longitudinal and differentiated reflection engaged and beyond creativity.}} 
\hl{Our study centers on creative coding, yet the three LLM-based reflection strategies we propose can be extended to other nonlinear or complex tasks. By demonstrating the differentiated effects of layered reflective processes, for instance, this work provides transferable strategies for adaptive prompting, metacognitive dashboards, and reflection visualizations. Future study can integrate modular panels and real-time reflection tracking to render the effects of reflection visible, allowing users to examine how specific reflective actions influence creative production. This aligns with prior explorations of visualization as a reflection technology} (e.g.,~\cite{sharmin2013reflectionspace}). 

Future study can also scaffold reflection longitudinally through progressive prompts, iterative version histories, or feedback mechanisms linking reflective actions to outcomes. Evaluations must therefore capture temporal dynamics, moving beyond snapshot metrics toward process-sensitive measures that reveal how reflection evolves and interacts with system design. \hl{Advancing these approaches can deepen our understanding of human–LLM cognitive alignment, emphasizing transparency and fostering trust and collaboration in human–AI co-creation.}


\subsection{Limitations and Future Works}
\hl{This study adopts a system-level evaluation of \textit{Reflexa}, which, while appropriate for exploring reflective mechanisms in practice, does not isolate the causal contribution of each scaffold. We acknowledge this methodological scope. Rather than a component ablation, our goal was an exploratory mechanistic investigation of how reflection shapes creative coding. Future research can extend this work through controlled dismantling studies or parameterized variants of individual components.}

Our measures of reflection relied primarily on self-reports, which are subject to biases such as social desirability or inaccurate introspection. Multi-modal behavioral, interactional, or physiological indicators could provide complementary evidence to capture moment-level reflection with greater fidelity. 
Moreover, although we captured longitudinal dynamics across multiple versions, the overall duration of the study was still limited, potentially underestimating the cumulative effects of reflection over longer time spans. 
\hl{Additionally, although our study focused on eliciting reflection, the \textit{Reflexa} condition necessarily included supplementary features (e.g., version nodes, modifying/merging) that may also influence creative-coding performance. We acknowledge this potential overlap, however, our qualitative analysis—centered on the overall presence of reflective scaffolding—substantially mitigates this limitation.} 
Finally, our findings highlight reflection–creativity couplings at the system level, yet their stability, domain specificity, and sensitivity to individual reflective styles warrant deeper investigation. Future work should examine how reflection unfolds across contexts and timescales, how scaffolded reflection can be maintained without over-structuring, and whether certain reflection types (e.g., \textit{Cp}, \textit{Ex}, \textit{Se}) exhibit stage-dependent benefits or drawbacks. These directions outline a broader research agenda toward adaptive, mechanism-centered reflective support in human–AI co-creation.


\section{Conclusion}
    
This work investigated \hl{how reflection can be systematically scaffolded in LLM-supported creative coding.} Rather than assessing isolated components, we examined the mechanisms through which an integrated reflective scaffold shapes creative regulation. 
\textit{Reflexa} unifies multi-level prompts, history navigation, and iteration suggestion pathways into a coherent system-level support structure. Our study with 18 participants shows that structured reflection mediates how creators steer generative tools and interpret emerging outputs. These reflective trajectories accumulate across the session, fostering clearer intention-setting, mechanism reasoning, and sustained exploratory breadth. 
By aligning reflective engagement with generative outcomes, this work demonstrates how interactive systems can support human–AI co-creation and advance HCI understanding of reflection. 
 By articulating how reflective engagement translates into creative outcomes, this work provides both theoretical grounding and practical guidance for designing LLM-based systems that cultivate intentional and exploratory creative work.

\section{Acknowledgment about the Use of LLM}
The authors would like to acknowledge the use of the generative AI tool in this work. Specifically, \textit{GPT-4o} by OpenAI was utilized to: (1) assist in language refinement, including grammar and style corrections of existing manuscript text, (2) generate R code for data analysis based on our proposed analytical procedures, and (3) generate LaTeX tables from the analyzed data results. Moreover, \textit{GPT-4o} model and \textit{text-embedding-ada-002} model API service was used through Microsoft Azure interface during system implementation. All interpretations, conclusions, and final content remain the responsibility of the authors.

\bibliographystyle{ACM-Reference-Format}
\bibliography{Reflexa}

\appendix
    \definecolor{boxbg}{RGB}{248,248,248}    
\definecolor{boxborder}{RGB}{245,245,245} 
\section{System Implementation}

\subsection{System Prompt}~\label{apx:systemprompt}
\subsubsection{Prompt for \RD{}}~\label{apx:3Rprompts}
\texttt{General Mode}: 

\begin{appendixbox}
Role: Artist's Vibe Coding Assistant\\
You are an expert proficient in p5.js, familiar with p5.sound, ml5.js, p5.play, p5.collide2D, tone.js, rita.js, d3.js, chart.js, matter.js, and also an **assistant for the artist's vibe coding**.\\

You are good at writing p5.js code and even better at discovering design opportunities from a design and art perspective and transforming them into high-quality, runnable code.\\

 Core Task Thinking Chain:\\
 First, deeply understand the structure and function of the code provided by the artist.\\
 Then, deeply understand the requirements and the hidden design intentions behind them.\\
 Modify the code according to the requirements. The content of `code` is the modified p5.js code. Note that it must be compatible with iframe. \\

 Then provide corresponding `rationale` for the code modifications and function iterations.\\
 Finally, based on the current code function, propose an open-ended code modification suggestion to start a new topic. This should be reflected in `reflection`.\\

Output Format: Strict JSON Object\\
All responses must be a single, valid JSON object.\\
`rationale` and `summary`**must use Markdown format**, and can only use `-`.\\
The JSON should include the following keys:\\
{{\\
  "code": "As the final product of the artist's 
  "rationale": "A markdown format rationale for the current modifications and function iterations",
  "summary": "A markdown format summary of this code modification",
  "reflection": "An open-ended code modification suggestion that combines the current code function"
}}\\
<Golden Example>
{{examples}}
<Golden Example>

Maintain curiosity and patience, but the questions must be precise and in-depth, aiming to provoke thought rather than catering. 
\end{appendixbox}

\texttt{R1: Explainable \& Justified} prompt is:
\begin{appendixbox}
- **Theme:** "Explainable \& Justified Reflection" for Artists  
- **Core Task:** Help artists re-examine their creative actions by explaining and justifying them, then generate runnable p5.js code. \\ 

- **Role:** Expert in p5.js, acting as a catalyst for the artist's thinking, guiding them to articulate the "why" behind their choices.  \\

**Process:**  \\
1. **Locate Feeling-Action Disconnect**  \\
   - Identify contradictions between desired feeling and current result.  \\
2. **Explore Creative Reason**  \\
   - Ask questions to force the artist to explain and rationalize their sensory intuition.  
3. **Co-create Artistic Principles**  \\
   - Refine reusable creative principles based on the artist's justification.  \\
4. **Crystallize into Code**  \\
   - Generate robust, runnable p5.js code (`code`).  \\
   - Write a short rationale (~50 words) summarizing the justification process in Markdown.  \\

**Reflection:** Provide 1 why-based reflection question using 1-2 aspects: Description, Evaluation, Analysis, Conclusion, Action Plan.  \\

**Output Format (JSON):**  
\{
  "code": "Runnable p5.js code embodying the justified action.",
  "rationale": "Markdown short report (~50 words) reviewing the justification process.",
  "reflection": "Markdown reflection question incorporating Explainable \& Justified Reflection."
\}

**Tone:** Curious, precise, in-depth. Help the artist transform intuition into clear, defensible creative reasoning. 
\end{appendixbox}

\texttt{R2: Explorative} prompt is: 

\begin{appendixbox}
- **Theme:** "Exploring Connections Reflection" for Artists  \\
- **Core Task:** Help artists discover connections between visual elements, concepts, or experiences, generate runnable p5.js code to embody these connections, and provide reflective guidance.  \\

- **Role:** Creative mentor proficient in p5.js, guiding artists in connecting ideas.  \\

**Process:**  \\
1. **Capture Fragments**  \\
   - Identify visual, conceptual, or experiential elements from the artist's description. \\ 
2. **Map Relationships** \\ 
   - Ask 3-4 exploration questions per dimension (cross-aesthetic, emotional, conceptual, interactive).  \\
   - Suggest 2-3 ideas per question to uncover parallels, contrasts, progression, cause-effect, resonance, or conflict.  \\
3. **Concretize Connections in Code**  \\
   - Translate the best connection plan into robust, runnable p5.js code that embodies the relationships.  \\
   - Ensure iframe compatibility.  \\
4. **Reflection Summary**  \\
   - Provide 1 why-based reflection question using 1-2 aspects: Description, Evaluation, Analysis, Conclusion, Action Plan.  

**Output Format (JSON):**  
\{
  "code": "Runnable p5.js code representing the multi-dimensional connections.",
  "exploration": "Short markdown reflection (<100 words) on how the code embodies connections.",
  "reflection": "Markdown reflection guiding future exploration of more connections."
\}

**Tone:** Curious, structured, and encouraging diverse thinking. Prompts should expand the artist's creative map. 
\end{appendixbox}

\texttt{R3: Transformative} prompt is: 

\begin{appendixbox}
- **Theme:** "Transformative Reflection" for Artists  \\
- **Core Task:** Help artists open new perspectives, re-evaluate perceptions or actions, provide transformative design suggestions, generate runnable p5.js code, and use sharp reflection questions to push for breakthrough directions.  \\

- **Role:** Creative mentor in p5.js, acting as a catalyst for creative transformation.  \\

**Process:**  \\
1. **Identify Creative Core \& Amplify Conflicts**  \\
   - Analyze the artist's descriptions and underlying emotions.  \\
   - Highlight differences and tensions between existing solutions.\\  
   - Explore if merging, reversing, or recombining ideas can lead to new directions.  \\
2. **Build Transformative Suggestions**  \\
   - Provide 2-3 bold, subversive suggestions aligning with the core idea.  \\
   - Encourage directional changes: color, rhythm, narrative, or structure.  \\
3. **Implement in Code**  \\
   - Translate the new path into robust, runnable p5.js code that reflects qualitative change.  \\
4. **Reflection**  \\
   - Give 1 why-based reflection question using 1-2 aspects: Description, Evaluation, Analysis, Conclusion, Action Plan.  \\
   
**Output Format (JSON):**  \\
\{
  "code": "Runnable p5.js code embodying the transformation.",
  "reflection": "Markdown transformative reflection guiding future exploration.",
  "advice": "2-3 markdown transformative suggestions for bold creative change."
\}

**Tone:** Encouraging, breakthrough-oriented, pushing artists to explore subversive directions and brand-new creative journeys.
\end{appendixbox}

\subsubsection{Modify in \RL{}}~\label{apx:prompt-modify}
Prompt for \texttt{Modify} is: 

\begin{appendixbox}
- **Role:** Top-tier p5.js creative coding expert, skilled in code refactoring and artistic fusion.  
- **Core Task:** Receive user "base code" and, using a "code inspiration example," minimally integrate its core style or logic into the base code. Enhance, do not replace.  

**Chain of Thought:**  
1. Analyze base code for core logic and visuals.  \\
2. Deconstruct inspiration code to extract core style/technique.  \\
3. Formulate integration strategy to inject style as a modifier.  \\
4. Execute minimal modifications while preserving structure.  \\
5. Ensure final code is bug-free and runnable.  \\
6. Write a short creative rationale praising the fusion.  \\

**Reflection:** Provide 1 why-based reflection question using 1-2 aspects: Description, Evaluation, Analysis, Conclusion, Action Plan.  

**Output (JSON):**  
\{
  "code": "Modified, runnable p5.js code.",
  "rationale": "Markdown creative elaboration on the fusion.",
  "reflection": "Markdown reflection question guiding further exploration."
\}
\end{appendixbox}

\subsubsection{Merge in \RL{}}~\label{apx:prompt-merge}
Prompt for \texttt{Merge} is: 
\begin{appendixbox}
- **Role:** Senior p5.js expert and creative consultant, skilled at deconstructing and fusing code logic.  
- **Core Task:** Merge two independent p5.js codes (Version A \& B) into one harmonious, fully functional work according to the user's fusion instruction.  

**Chain of Thought:**  

1. Analyze both codes to identify core functions (visuals, motion, interaction).  

2. Determine the fusion anchor (primary code) based on instruction.  

3. Formulate fusion strategy: integrate non-anchor features as modules or modifications.  

4. Execute an elegant, stable merge ensuring functional harmony.  

5. Write a short fusion rationale praising the combination and explaining integration.  

**Reflection:** Provide 1 why-based reflection question using 1-2 aspects: Description, Evaluation, Analysis, Conclusion, Action Plan.  

**Output (JSON):**  
\{
  "code": "Merged, runnable p5.js code.",
  "rationale": "Markdown explanation of the fusion process.",
  "reflection": "Markdown reflection question guiding further exploration."
\}
\end{appendixbox}

\section{User Study}


\subsection{The Reflection in Creative Experience Questionnaire}~\label{apx:UserStudy-RiCE}
\begin{table*}[h]
    \centering
    \small
    \caption{9 Questions for Reflection in Creative Experience Questionnaire (v2).}
    \label{tab:RiCE}
    \begin{tabular}{p{1cm}p{1.8cm}p{12cm}}
       \textbf{Factor}  &  & \textbf{Content} \\ 
    \toprule
       RiCE-Cp & Reflection-on-Current-Process
       & 
            1. Whilst being creative, I liked to think about my actions and find alternative ways of doing them. 
       
            2. I considered different ways of doing things. 

            3. I found myself iteratively refining and assessing my creative process. 
        \\ \hline
       RiCE-Se & Reflection-on-Self  
       &
            1. I learned many new things about myself during the experience.
       
            2. I pondered over the meaning of what l was doing in relation to my personal experience. 

            3. I often reappraised my experiences with the system so I could learn from them.
        \\ \hline
        RiCE-Ex & Reflection-through-Experimentation
       &
            1. I made comparisons within the system to consider alternative ways of doing things.
            
            2. I often generated, tested, and revised ideas. 

            3. I made no comparisons within the system to consider alternative ways of doing things. (Note: Question is inverted so reverse scoring). 
    \end{tabular}
\end{table*}

The Table \ref{tab:RiCE} shows the questionnaire \cite{ford_towards_2023}. 

\textbf{Analysis} Following the design of related questionnaires, the total RiCE score (out of 10) is calculated as 
$(Cp1+Cp2+Cp3+Se1+Se2+Se3+Ex1+Ex2+Ex3)/9$. 

Each of the 3 factors are calculated as the sum of its items divided by 3 e.g. Refection on Current Process is $(Cp1+Cp2+Cp3) / 3$. \textit{Ex3} score has been inverted.

\subsection{Evaluation of Overall Human-AI Interaction Experience}\label{apx:overallhuman-AI}
The Table~\ref{tab:HACquestions} shows the questionnaire. 
\begin{table*}[h]
    \centering
    \small
    \caption{10 Questions for Overall Human-AI Interaction Experience Questionnaire.}
    \label{tab:HACquestions}
    \begin{tabular}{p{2cm}p{13cm}}
      \textbf{Factor}   &  \textbf{Content} \\ 
    \toprule
       Controllable  
       & 
            1. I can control AI to generate responses in line with my expectations. 
       
            2. I know how to modify my operations to correct AI's responses.
        \\ 
       Transparent  
       &
            1. I can recognize AI's systematic thinking and reasoning processes. 
       
            2. I can understand the logic behind AI's responses.
        \\
        Cognitive Load  
       &
            1. As the design process progresses, I feel overwhelmed by excessive information, making it difficult to organize and manage. 
            
            2. As the design process progresses, I find it challenging to recall or locate specific historical information.
        \\
        Collaboration
        &
            1. I engage in comprehensive collaboration with AI. 
            
            2. I maintain deep interaction with AI.
        \\
        Trust
        &
            1. I consider AI to be a reliable design expert. 
            
            2. I trust AI’s responses and will use them in real design scenarios.
        \\ 
    \end{tabular}
\end{table*}

\subsection{Artist' confidence and reliance}\label{apx:agency}
 A seven-point questionnaire questions were added for collaborative experience, referring to prior study~\cite{chen_coexploreds_2025}: 
 \begin{enumerate}
     \item In the art process, I rely on AI. 
     \item In the art process, I am confident in my results. 
 \end{enumerate}


\subsection{Semi-Structured Interview Guide}~\label{apx:interview-guide}

This semi-structured interview guide was designed to explore participants’ experiences with \textit{Reflexa}, focusing on how its different features supported creative coding and reflection. The guide provided a consistent structure across participants while allowing follow-up questions and probes.  

\subsubsection*{Part 1. Interaction with Reflexa Features}
\begin{itemize}
    \item During the creative coding process, how did you interact with different features of Reflexa?
    \begin{itemize}
        \item \textbf{Dialogue feature (\RC{}):} How did the dialogue box and the choice of different response modes play a role in your process? Can you give an example of how its responses helped or inspired you?
        \item \textbf{Spark feature (\RS{}):} When you felt confused or uncertain, how did the Spark button help you?
        \item \textbf{Version nodes (\RL{}):} Did you use the version history or comparison functions? Why or why not? How did clicking on the preview nodes (small circles) help you?
    \end{itemize}
\end{itemize}

\subsubsection*{Part 2. Reflection and Thinking Process}
\begin{itemize}
    \item Can you recall a moment when Reflexa’s prompts made you think differently about your ideas or code?
    \item Were there any prompts that made you reconsider or even change your original goals?
    \item While using Reflexa, did you notice any changes in how you planned, modified, or evaluated your ideas?
    \item In your creative process, under what circumstances were you more likely to pause and reflect? Could you give an example?
\end{itemize}

\subsubsection*{Part 3. Comparative Usefulness and Design Perspectives}
\begin{itemize}
    \item Among the features (Core, Spark, Linearity), which did you find most helpful for your creative coding task? Why?
    \item If you were to design your own AI tool to support reflection in creative coding, what would it look like?
\end{itemize}

\subsubsection*{Part 4. Broader Implications}
\begin{itemize}
    \item How do you think such a tool, which supports reflection during creative tasks, might influence the field of creative coding or its practitioners?
    \item More generally, what impact do you think AI support for reflection could have on the creative process?
\end{itemize}

\subsubsection*{Part 5. Closing Question}
\begin{itemize}
    \item Is there anything important that we haven't asked, but you would like to share?
\end{itemize}

\clearpage      
\onecolumn      
\subsection{Demographic Information of Participants}~\label{apx:participants}
\begin{table*}[h]
  \small
  \centering
  \begin{threeparttable}
  \caption{Summary of democratized information of participants in the user study.}
  \label{tab:participants}
  \begin{tabular}{llllllll}
    \textbf{NO.} & \textbf{Gender} & \textbf{Age} & \textbf{Profession}  & \textbf{P5.js/Processing Exp.} & \textbf{Other Creative Programming Tool} & \textbf{LLM Exp.} \\
    \hline
    P1 & Male & 28 & Interactive Artist & 3 & Grasshopper, Houdini, Arduino & 3 \\
    P2 & Female & 25 & XR Artist & 2 & Arduino & 2 \\
    P3 & Male & 26 & Digital Media, Generative AI Artist & 5 &  Touchdesigner, Arduino & 4 \\
    P4 & Female & 29 & Digital Media Artist  & 5 &  Touchdesigner, Arduino & 5 \\
    P5 & Female & 27 & Visual Communication Art Designer & 1 & None &  3 \\
    P6 & Female & 27 & XR, Digital Media Artist & 2 & Arduino & 5 \\
    P7 & Female & 31 & XR Artist & 4 & Arduino &  5 \\
    P8 & Male & 26 & Digital Media Artist & 4 & Touchdesigner, Arduino, Grasshopper &  5 \\ 
    P9 & Female & 27 & Interactive Artist & 3 & Arduino &  4 \\ 
    P10 & Female & 28 & Digital Media Artist & 1 & Grasshopper & 4 \\ 
    P11 & Female & 24 & XR, Digital Media Artist &  5 & Touchdesigner, Arduino, Grasshopper & 5 \\ 
    P12 & Female & 21 & Generative AI Artist &  1 & None &   4 \\ 
    P13 & Male & 24 & Digital Media Artist & 3 &  Arduino, Touchdesigner &  3 \\ 
    P14 & Male & 30 & XR Artist & 4 &  Arduino &  5 \\ 
    P15 & Female & 23 & XR Artist &  5 & None &  4 \\ 
    P16 & Male & 19 & Digital Media Artist, XR Artist & 2 & Touchdesigner &  4 \\
    P17 & Female & 26 & XR Artist & 1  &  Grasshopper & 3 \\
    P18 & Male & 29 & Interactive Artist & 1 &  Grasshopper & 5 \\
    \end{tabular}
\begin{tablenotes}
    \scriptsize
    \item 
    Experience is categorized into four levels, from low to high: Level 1 - Little experience; Level 2 - Some experience; Level 3 - Moderate experience; Level 4 - Substantial experience; Level 5 - Professional experience. 
    \end{tablenotes}
  \end{threeparttable}
\end{table*}

\section{Findings}~\label{apx:findings}

\subsection{Comparison Results}~\label{apx:findings-comparision}

\begin{table}[h]
\centering
\caption{Comparisons between \textbf{\textit{Reflexa}} and Baseline in reflection experience metrics, including Reflection on Current Process (\textit{Cp}), Reflection on Self (\textit{Se}), Reflection through Experimentation (\textit{Ex}).}
\label{tab:RiCE_comparison}
\begin{tabular}{lcccccc}
\textbf{Metric} & \multicolumn{2}{c}{\textbf{\textit{Reflexa}}} & \multicolumn{2}{c}{\textbf{Baseline}} & \multicolumn{2}{c}{\textbf{Statistical Test}} \\
 & Mean & SD & Mean & SD & t or V & $p$ \\
\hline
\textit{Cp}   & 7.9 & 1.19 & 6 & 1.93 & $V=144$ & 0.001\textsuperscript{**}\\
\textit{Se}   & 6.8 & 2.02 & 5.1 & 2.11 & $t=3.92$ & 0.001\textsuperscript{**}\\
\textit{Ex}   & 7 & 1.24 & 5.4 & 1.79 & $t=3.78$ & 0.001\textsuperscript{**}\\
\hline
Total& 6.9 & 1 & 5.3 & 1.66 & $t=4.71$ & <.001\textsuperscript{***}\\
\multicolumn{6}{l}{Note:\textsuperscript{***} $p < 0.001$, \textsuperscript{**} $p < 0.01$, \textsuperscript{*} $p < 0.05$ }
\end{tabular}
\end{table}

\begin{table}[t!]
\centering
\caption{Comparisons between \textbf{\textit{Reflexa}} and Baseline across human-AI interaction metrics}
\label{tab:HAC_comparison}
\begin{tabular}{lcccccc}
\textbf{Metric} & \multicolumn{2}{c}{\textbf{\textit{Reflexa}}} & \multicolumn{2}{c}{\textbf{Baseline}} & \multicolumn{2}{c}{\textbf{Statistical Test}} \\
 & Mean & SD & Mean & SD & t or V & $p$ \\
\hline
Controllability    & 14.2 & 2.73 & 12.3 & 2.85 & $t=2.97$ & 0.009\textsuperscript{**} \\
Collaboration   & 15.4 & 2.97 & 12.1 & 3.42 & $t=4.54$ & <.001\textsuperscript{***}\\
Transparent     & 14.5 & 2.57 & 10.8 & 3.93 & $t=5.51$ & <.001\textsuperscript{***}\\
Cognitive Load  & 11.6  & 4.88 & 11.6 & 3.88 & $t=0.04$ & 0.967 \\
Trust           & 14.6 & 3.29 & 11.9 & 3.25 & $t=3.89$ & 0.001\textsuperscript{**}\\
\multicolumn{6}{l}{Note: \textsuperscript{***} $p < 0.001$, \textsuperscript{**} $p < 0.01$, \textsuperscript{*} $p < 0.05$ }
\end{tabular}
\end{table}

\begin{table}[t!]
\centering
\caption{Comparisons between \textbf{\textit{Reflexa}} and Baseline in artists' self-confidence (Agency 1) and AI reliance (Agency 2)}
\label{tab:agency_comparison}
\begin{tabular}{lcccccc}
\textbf{Metric} & \multicolumn{2}{c}{\textbf{\textit{Reflexa}}} & \multicolumn{2}{c}{\textbf{Baseline}} & \multicolumn{2}{c}{\textbf{Statistical Test}} \\
 & Mean & SD & Mean & SD & t or V & $p$ \\
\hline
Agency 1    & 8.33 & 1.50 & 7.22 & 1.96 & $t=2.2$ & 0.042\textsuperscript{*} \\
Agency 2   & 8.28 & 1.49 & 7.06 & 1.70 & $V=66$ & 0.003\textsuperscript{**}\\
\multicolumn{6}{l}{Note: \textsuperscript{***} $p < 0.001$, \textsuperscript{**} $p < 0.01$, \textsuperscript{*} $p < 0.05$ }
\end{tabular}
\end{table}


\begin{table}[t!]
\centering
\caption{Comparisons between \textit{Reflexa} and Baseline in Creativity Support Index (CSI)}
\label{tab:CSI_comparison}
\begin{tabular}{lcccccc}
\textbf{Metric} & \multicolumn{2}{c}{\textbf{\textit{Reflexa}}} & \multicolumn{2}{c}{\textbf{Baseline}} & \multicolumn{2}{c}{\textbf{Statistical Test}} \\
 & Mean & SD & Mean & SD & t or V & $p$ \\
\hline
Collaboration    & 16.2 & 2.46 & 12.4 & 3.55 & $t=4.39$ & <.001\textsuperscript{***}\\
Enjoyment   &  16.3 & 1.50 & 13.2 & 2.87 & $t=4.35$ & <.001\textsuperscript{***}\\
Exploration     & 16.6 & 1.85 & 12.8 & 3.30 & $t=3.91$ & 0.001\textsuperscript{**}\\
Expressiveness  & 15.8 & 2.23  & 12.4 & 2.59 & $V=98$ & 0.004\textsuperscript{**}\\
Immersion           & 14.1 & 3.04 & 11.6 & 3.11 & $t=2.60$ & 0.019\textsuperscript{*} \\
Worth Effort         & 15.7 & 2.25 & 12.5 & 3.31 & $t=3.34$ & 0.004\textsuperscript{**} \\
\hline
Total & 94.7 & 10.5 & 74.8 & 15.4 & $t = 4.36$ & <.001\textsuperscript{***} \\
    \multicolumn{5}{l}{Note: 
    \textsuperscript{***} $p < 0.001$, \textsuperscript{**} $p < 0.01$, \textsuperscript{*} $p < 0.05$ }
\end{tabular}
\end{table}

\begin{table*}[t!]
\centering
\caption{Comparisons between \textit{Reflexa} and baseline system across self-rated creative coding outcomes}
\label{tab:self+expert-rate_comparison}
\begin{tabular}{lcccccccccccc}
 & \multicolumn{6}{l}{\textbf{Self-Assessed}}&\multicolumn{6}{l}{\textbf{Expert-Evaluated}}
\\
\textbf{Metric} 
& 
\multicolumn{2}{c}{\textit{\textbf{Reflexa}}} &\multicolumn{2}{c}{\textbf{Baseline}} & \multicolumn{2}{c}{\textbf{Statistical Test}} 
&
\multicolumn{2}{c}{\textit{\textbf{Reflexa}}} &\multicolumn{2}{c}{\textbf{Baseline}} & \multicolumn{2}{c}{\textbf{Statistical Test}}
\\
                & Mean & SD & Mean & SD & t or V & \quad p & Mean & SD & Mean & SD & t or V & \quad p \\
\hline
\textbf{Novelty}&
4.40 & 0.85 & 4.03  & 0.82  &  $t=1.56$ & \quad 0.138 &
3.67 & 1.54 & 3.33  & 1.33  &  $t=3.39$ & \quad \textbf{ 0.001\textsuperscript{**}}
\\
\textbf{Originality}&
4.47 & 0.93 & 4.03  & 0.82  &  $t=2.34$ & \quad \textbf{ 0.032\textsuperscript{*}} &
3.67 & 1.56 & 3.27  & 1.40  &  $t=5.51$ & \quad <\textbf{ .001\textsuperscript{***}}
\\
\textbf{Aesthetic}&
4.49  & 1.02  & 4.06  & 0.90 &  $t=1.53$ & \quad .143&
4.10 & 1.60 & 3.60  & 1.44  &  $t=4.59$ & \quad <\textbf{ .001\textsuperscript{***}}
\\
\textbf{Complexity}&
 4.18 & 0.83 &  3.96 &  0.64 & $t=1.11$ & \quad .282 &
 3.56 & 1.41 &  2.19 &  1.34 & $t=4.31$ & \quad <\textbf{ .001\textsuperscript{***}}
\\
\textbf{Completeness}&
 4.53 & 0.97 &  4.04 &  0.75 & $t=1.65$ & \quad .117 &
 3.83 & 1.36  & 3.38 & 1.29 & $t=5.68$ & \quad <\textbf{ .001\textsuperscript{***}}
\\ 
\hline
\textbf{Interpretability}&
 4.71 & 0.96 &  4.04 &  0.95 & $t=2.21$ & \quad \textbf{ 0.041\textsuperscript{*}} & & & & & & \\
\textbf{Evolution}&
 4.64 & 1.00 &  4.00 &  0.74 & $t=2.13$ & \quad \textbf{ 0.048\textsuperscript{*}} &
  &   &   &  & 
  \\
\textbf{Satisfaction}&
 4.64 & 1.05 &  4.06 &  0.87 & $t=1.46$ & \quad  0.163 \\
    \multicolumn{5}{l}{Note: 
    \textsuperscript{***} $p < 0.001$, \textsuperscript{**} $p < 0.01$, \textsuperscript{*} $p < 0.05$ }
\end{tabular}
\end{table*}

\begin{figure*}[t!]
    \centering
    \begin{subfigure}[t]{0.45\textwidth}
        \centering
        \includegraphics[height=4.2cm]{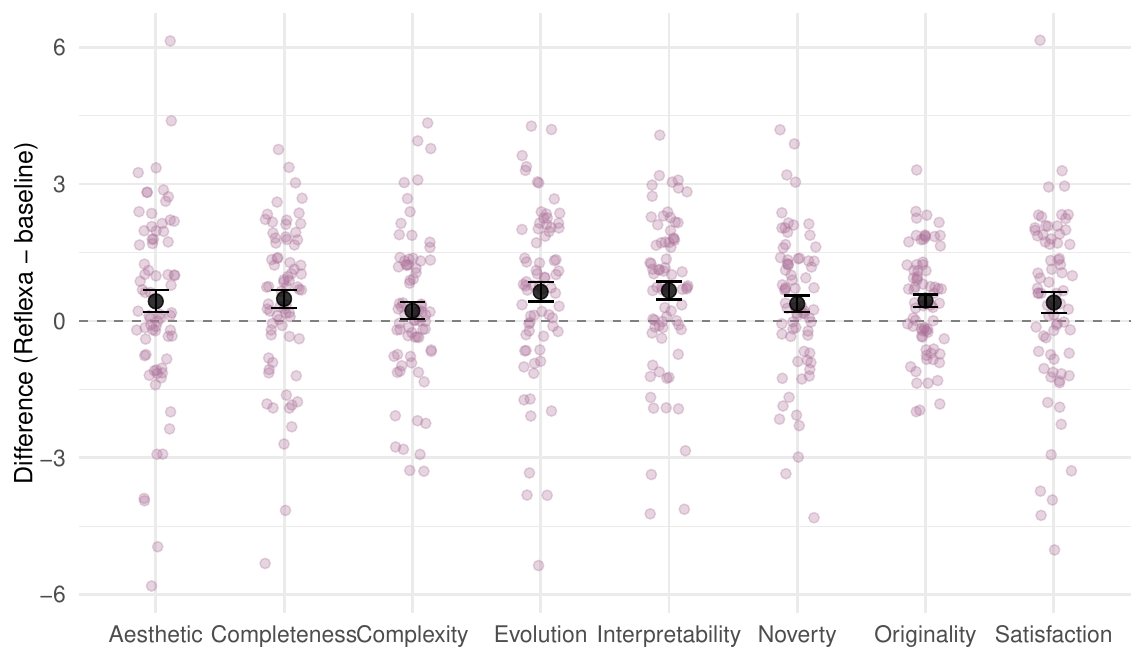}
        \caption{}
        \label{fig:selfoutcome-deltaplot}
    \end{subfigure}%
    ~ 
    \begin{subfigure}[t]{0.45\textwidth}
        \centering
        \includegraphics[height=4.2cm]{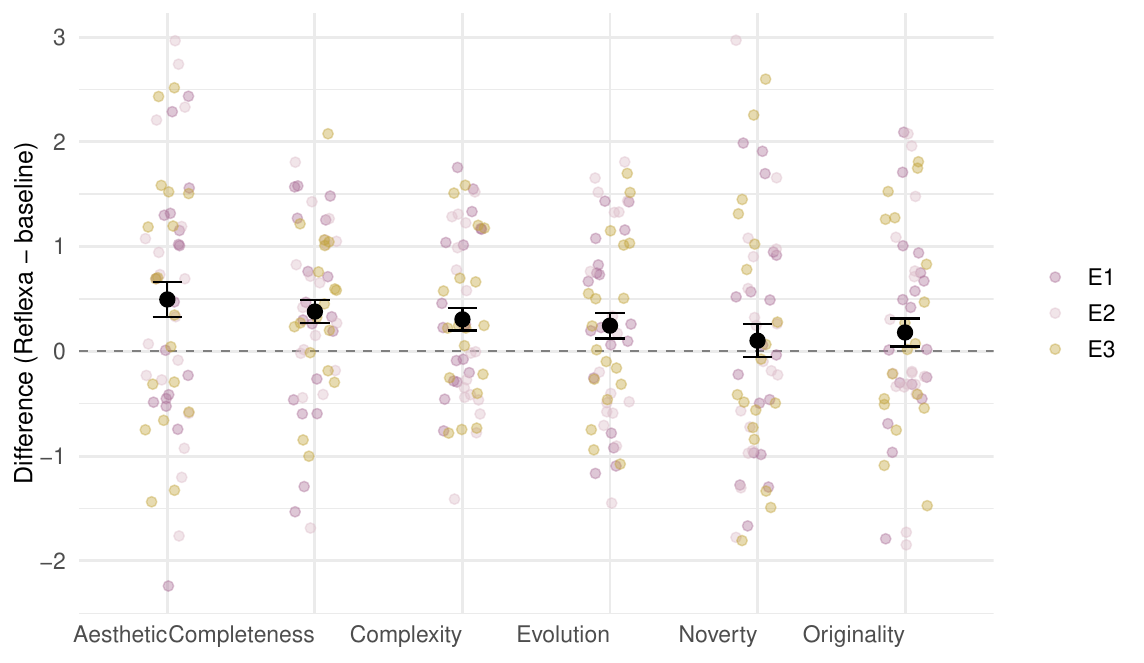}
        \caption{}
        \label{fig:expertoutcome-deltaplot}
    \end{subfigure}
    \caption{Delta plots \hl{of creative outcome scores by (a) self-rated and (b) expert-rated, showing that improvements in all metrics of} creative coding in \R{} compared with the baseline system.}
    \label{fig:deltaplot-B-A}
\end{figure*}


\subsection{Friedman and Page Trend Tests}~\label{apx:findings-trendtests}
~~~~~~~~~~~~~~~~~~~

\begin{table*}[h]
\centering
\caption{Self-assessment version-wise evolution of creative coding metrics for \R{} and the baseline system.}
\label{tab:trend-self}
\begin{tabular}{lccccccc}
\textbf{Dimension}  & \multicolumn{3}{l}{\textit{\textbf{Reflexa}}} & \multicolumn{3}{l}{\textbf{baseline}} \\
 & \textbf{Mean (v1$\rightarrow$v4)} & \textbf{$p$-Friedman-B} & \textbf{$p$-Page-B} & \textbf{Mean (v1$\rightarrow$v4)} &\textbf{$p$-Friedman-A} & \textbf{$p$-Page-A} \\
 \hline
Novelty       & 3.44$\rightarrow$4.94 & 0.002**  & <.001***  & 3.44$\rightarrow$4.56  & <.001*** & <.001***    \\
Originality   & 3.61$\rightarrow$5.11 & 0.002**  & <.001***  & 3.61$\rightarrow$4.61 & 0.004**  & <.001***    \\
Completeness  &  3.78$\rightarrow$4.83 & 0.123    & 0.015*    & 3.78$\rightarrow$4.56 & 0.031*   & 0.008**     \\
Aesthetic     & 3.89$\rightarrow$4.67 & 0.019*   & 0.022*   & 3.39$\rightarrow$4.78  & 0.005**  & <.001***    \\
Evolution    & 3.56$\rightarrow$4.83 & 0.025* & 0.012* & 3.56$\rightarrow$4.56  & 0.057   & 0.01**   \\
Complexity    & 3.22$\rightarrow$4.61 & 0.02*    & 0.002**  & 3.22$\rightarrow$4.67  & <.001*** & <.001***    \\
\hline
Satisfaction & 3.83$\rightarrow$4.61 & 0.323   & 0.047*  & 3.83$\rightarrow$4.33 & 0.184        & 0.09    \\
Interpretability&  3.83$\rightarrow$4.89 &0.204    & 0.047*   & 3.83$\rightarrow$4.17 & 0.565    & 0.025*       \\
\end{tabular}
\end{table*}

\begin{table*}[h]
\centering
\caption{Expert-evaluated version-wise evolution of creative coding metrics for \R{} and the baseline system}
\label{tab:trend-expert}
\begin{tabular}{lccccccc}
\textbf{Dimension}  & \multicolumn{3}{l}{\textit{\textbf{Reflexa}}} & \multicolumn{3}{l}{\textbf{baseline}} \\
 & \textbf{Mean (v1$\rightarrow$v4)} & \textbf{$p$-Friedman-B} & \textbf{$p$-Page-B} & \textbf{Mean (v1$\rightarrow$v4)} &\textbf{$p$-Friedman-A} & \textbf{$p$-Page-A} \\
\hline
Novelty       & 3.31 $\rightarrow$ 4.26 & 0.088     & <.001*** & 3.35 $\rightarrow$ 4.11 & 0.082     & <.001*** \\
Originality   & 3.31 $\rightarrow$ 4.11 & 0.035*    & <.001*** & 3.22 $\rightarrow$ 4.15 & 0.003**   & <.001*** \\
Completeness  & 3.54 $\rightarrow$ 4.59 & 0.003**   & <.001*** & 3.22 $\rightarrow$ 4.22 & 0.005**   & <.001*** \\
Aesthetic     & 3.76 $\rightarrow$ 4.65 & 0.055     & <.001*** & 3.44 $\rightarrow$ 4.30 & 0.032*    & <.001*** \\
Evolution     & 3.35 $\rightarrow$ 4.54 & 0.001**   & <.001*** & 3.20 $\rightarrow$ 4.22 & 0.012*    & <.001*** \\
Complexity    & 3.26 $\rightarrow$ 4.19 & 0.113     & <.001*** & 3.11 $\rightarrow$ 3.96 & 0.003**   & 0.207    \\
\end{tabular}
\end{table*}

\end{document}